\documentclass[twocolumn]{aastex631}

\usepackage{array}
\usepackage{makecell}
\usepackage{hyperref}
\usepackage{multirow}
\usepackage{makecell}
\usepackage{rotating}

\newcommand{\ebvgas}{$E(B-V)_{\text{gas}}$}
\newcommand{\one}{~\textsc{i}}
\newcommand{\ii}{~\textsc{ii}}
\newcommand{\iii}{~\textsc{iii}}
\newcommand{\iv}{~\textsc{iv}}

\newcommand{\mstar}{M$_*$}

\newcommand{\te}{$T_{\mathrm{e}}$}
\newcommand{\den}{$n_{\mathrm{e}}$}
\newcommand{\oiiia}{[O~\textsc{iii}]$\lambda4364$}
\newcommand{\oiiis}{[O~\textsc{iii}]$\lambda5008$}
\newcommand{\oiia}{[O~\textsc{ii}]$\lambda\lambda$7322,7332}

\shorttitle{High-redshift metallicity calibrations}
\shortauthors{Sanders et al.}

\begin{document}

\title{Direct $T_{\mathrm{e}}$-based Metallicities of $z=2-9$ Galaxies with {\it JWST}/NIRSpec: Empirical Metallicity Calibrations Applicable from Reionization to Cosmic Noon}

\author[0000-0003-4792-9119]{Ryan L. Sanders}\altaffiliation{NHFP Hubble Fellow}\affiliation{Department of Physics and Astronomy, University of Kentucky, 505 Rose Street, Lexington, KY 40506, USA}\affiliation{Department of Physics and Astronomy, University of California, Davis, One Shields Ave, Davis, CA 95616, USA}

\email{email: ryan.sanders@uky.edu}

\author[0000-0003-3509-4855]{Alice E. Shapley}\affiliation{Department of Physics \& Astronomy, University of California, Los Angeles, 430 Portola Plaza, Los Angeles, CA 90095, USA}

\author[0000-0001-8426-1141]{Michael W. Topping}\affiliation{Steward Observatory, University of Arizona, 933 N Cherry Avenue, Tucson, AZ 85721, USA}

\author[0000-0001-9687-4973]{Naveen A. Reddy}\affiliation{Department of Physics \& Astronomy, University of California, Riverside, 900 University Avenue, Riverside, CA 92521, USA}

\author[0000-0003-2680-005X]{Gabriel B. Brammer}\affiliation{Cosmic Dawn Center (DAWN)}\affiliation{Niels Bohr Institute, University of Copenhagen, Jagtvej 128, 2200 Copenhagen N, Denmark}

\begin{abstract}
We report detections of the [O\iii]$\lambda$4364 auroral emission line for 16 galaxies at $z=2.1-8.7$, measured from {\it JWST}/NIRSpec observations obtained as part of the Cosmic Evolution Early Release Science (CEERS) survey program.
We combine this CEERS sample with 9 objects from the literature at $z=4-9$ with auroral-line detections from {\it JWST}/NIRSpec and 21 galaxies at $z=1.4-3.7$ with auroral-line detections from ground-based spectroscopy.
We derive electron temperature (\te) and direct-method oxygen abundances for the combined sample of 46 star-forming galaxies at $z=1.4-8.7$.
We use these measurements to construct the first high-redshift empirical \te-based metallicity calibrations for the strong-line ratios [O\iii]/H$\beta$, [O\ii]/H$\beta$, R23=([O\iii]+[O\ii])/H$\beta$, [O\iii]/[O\ii], and [Ne\iii]/[O\ii].
These new calibrations are valid over 12+log(O/H$)=7.4-8.3$ and can be applied to samples of star-forming galaxies at $z=2-9$, leading to an improvement in the accuracy of metallicity determinations at Cosmic Noon and in the Epoch of Reionization.
The high-redshift strong-line relations are offset from calibrations based on typical $z\sim0$ galaxies or H\ii\ regions, reflecting the known evolution of ionization conditions between $z\sim0$ and $z\sim2$.
Deep spectroscopic programs with {\it JWST}/NIRSpec promise to improve statistics at the low and high ends of the metallicity range covered by the current sample, as well as improve the detection rate of [N\ii]$\lambda$6585 to allow the future assessment of N-based indicators.
These new high-redshift calibrations will enable accurate characterizations of metallicity scaling relations at high redshift, improving our understanding of feedback and baryon cycling in the early universe.
\end{abstract}

\section{Introduction} \label{sec:intro}

The abundance of heavy elements relative to hydrogen, or metallicity, is a fundamental property of galaxies that traces the combined effects of stellar mass buildup and gas flows that add or remove mass and metals from systems.
Theoretical models of galaxy evolution describe how the gas-phase metallicity of the interstellar medium (ISM), traced by the oxygen abundance O/H, is set by the relative strength of the star-formation rate (SFR), mass inflow rate, and mass outflow rate \citep[e.g.,][]{dav2012,lil2013,tor2019}.
A major goal of modern astronomy is thus to robustly characterize the way metallicity scales with galaxy properties including stellar mass (\mstar) and SFR and how such scaling relations change with redshift to understand gas flows and baryonic mass assembly across cosmic history.

In the local universe, gas-phase metallicity increases with increasing \mstar, following a tight mass-metallicity relation \citep[MZR; e.g.,][]{leq1979,tre2004,kew2008,ber2012,and2013,cur2020}.
A three-dimensional ``fundamental metallicity relation'' (FMR) between metallicity, \mstar, and SFR has also been identified in which metallicity decreases with increasing SFR at fixed \mstar\ \citep[e.g.,][]{ell2008,man2010,lar2010,and2013,cre2019,cur2020}.
The MZR and FMR have been extensively characterized at $z\sim0$.

Sensitive near-infrared spectrographs on large ground-based telescopes and {\it HST} have provided measurements of rest-optical line ratios at $z\sim1-4$, enabling metallicity studies to be carried out in the first half of cosmic history.
Such studies have generally found that metallicity decreases at fixed \mstar\ with increasing redshift \citep[e.g.,][]{erb2006,san2015,san2021,pap2022,str2022} and that the FMR does not strongly evolve out to $z\sim3.5$ \citep{cre2019,san2020,san2021}.
While spectroscopic metallicity samples at $z\sim2-3$ now comprise hundreds of galaxies such that the statistical precision is high, there is significant systematic uncertainty on derived metallicities due to the unknown form of the relations between rest-optical strong-line ratios and O/H at high redshift.
The lack of robust high-redshift metallicity calibrations likewise limits conclusions drawn from the fast-growing archive of rest-optical spectra from {\it JWST} for galaxies at $z>4$ and reaching deep into the epoch of reionization \citep[e.g.,][]{sha2023b,san2023bpt,nak2023,mat2022,bun2023}.

Empirical calibrations between rest-optical line ratios and metallicity can be constructed using samples for which O/H has been derived using the robust ``direct method'' that is based on electron temperature (\te) determinations.
In this approach, \te\ is calculated from the flux ratio of a faint auroral emission line (e.g., \oiiia) to a bright line from the same ion (\oiiis), leveraging the fact that the two transitions arise from different upper energy levels.
This temperature can then be used to calculate the emissivity of various transitions to convert dust-corrected flux ratios of O ionic lines to H recombination lines (i.e., [O\iii]/H$\beta$ and [O\ii]/H$\beta$) into O/H.
Metallicity calibrations can then be constructed by fitting functional forms to the relations between different line ratios and direct-method O/H.
This approach has been used to construct many different metallicity calibrations based on local H\ii\ regions and $z\sim0$ star-forming galaxies that form the basis of MZR and FMR studies carried out on large samples\citep[e.g.,][]{pet2004,mai2008,mar2013,cur2017,cur2020}.

There is now significant evidence that, at fixed O/H, the ionization conditions of the ISM evolve from ``normal'' conditions at $z\sim0$ toward a more extreme state at $z\sim2$ associated with a harder ionizing spectrum due to the $\alpha$-enhanced chemical abundance patterns of young stars and elevated electron densities \citep[e.g.,][]{ste2014,ste2016,san2016,san2020,str2017,str2018,sha2015,sha2019,run2021,top2020a,top2020b,cul2021}.
Since the ionization conditions set the shape of metallicity calibrations, an important consequence of these results is that metallicity calibrations are expected to change between $z\sim0$ and $z\sim2$.
Accordingly, calibrations based on typical $z\sim0$ sources should not be applied to high-redshift samples.
To address this issue, \te-based calibrations were constructed using low-redshift galaxies with extreme line-ratio or SFR properties similar to those of typical high-redshift samples, assuming that matching in such properties selects sources with the same ISM ionization conditions present at high redshift \citep{bia2018,per2021,nak2022}.
However, the validity of these ``analog'' calibrations must ultimately be tested directly at high redshift.

The most robust resolution to this problem is to construct high-redshift metallicity calibrations using direct metallicity and strong-line measurements of high-redshift galaxies themselves.
Based on deep ground-based spectroscopy of bright high-redshift line emitters, a sample of $\sim20$ star-forming galaxies at $z=1.4-3.7$ with auroral line detections and direct-method metallicities has been assembled, representing many nights of $8-10$~meter telescope time \citep{vil2004,bra2012b,chr2012a,chr2012b,sta2013,sta2014,bay2014,jam2014,san2016b,koj2017,ber2018,pat2018,gbu2019,gbu2022,san2020,san2023oiia}.
\citet{san2020} showed that, on average, this sample deviates from calibrations based on normal $z\sim0$ sources, but is well-matched by the local analog calibrations of \citet{bia2018}.
However, both the sample size and the fidelity of individual measurements of this ground-based sample fall short of what is required to construct new calibrations.
As discussed in \citet{san2023oiia}, this shortcoming is primarily due to the sensitivity provided by current ground-based near-infrared spectrographs, the highly wavelength-dependent sky background, and the limited accessible near-infrared wavelength ranges due to atmospheric transmission.
Accordingly, a significant expansion of the $z>1$ auroral-line sample is not feasible with current ground-based facilities.

{\it JWST} now provides the spectroscopic capabilities to overcome all of the challenges described above and obtain a large sample of high-redshift galaxies with direct-method metallicities for the first time, paving the way toward the first robust high-redshift metallicity calibrations.
Upon commencing science operations, the NIRSpec instrument onboard {\it JWST} immediately demonstrated the ability to detect auroral emission lines of distant galaxies in the Early Release Observations (ERO) targeting the SMACS~0723 cluster field.
The ERO spectra revealed detections of \oiiia\ for three galaxies at $z>7.5$ for which \te-based metallicities were reported \citep{cur2023,bri2022,sch2022,are2022,tay2022,tru2022,tac2022}.
NIRSpec observations from the Cosmic Evolution Early Release Science (CEERS) and GLASS Early Release Science (ERS) programs have yielded direct-method metallicities for several additional sources at $z=4-9$ \citep{tay2022,nak2023,tang2023,jon2023}.
However, these early studies were limited to a small number of {\it JWST} targets and furthermore did not integrate all of the existing ground-based \te\ data at $z\sim1-4$.


In this paper, we report detections of \te-sensitive auroral emission lines for 16 galaxies at $z=2.1-8.7$ measured from medium-resolution {\it JWST}/NIRSpec spectroscopy from the CEERS survey, which we use to derive robust gas-phase oxygen abundances using the direct method.
Of these detections, 11 are new while 5 have been previously reported \citep{nak2023,tang2023}.
We combine this sample with 9 sources at $z=4-8.5$ from the literature with detections of auroral lines from other {\it JWST} ERO and ERS programs and 21 targets at $z=1.4-3.6$ drawn from the literature with direct-method metallicities from ground-based spectroscopy.
We use the resulting sample of 46 galaxies at $z=1-9$ to derive the first empirical high-redshift metallicity calibrations that enable a robust translation of rest-optical strong-line ratios into gas-phase oxygen abundance.
These relations are valid from Cosmic Noon into the Epoch of Reionization, and over the metallicity range 12+log(O/H$)=7.4-8.3$
This paper is organized as follows. 
We describe the observations, data reduction, and measurements in Section~\ref{sec:obs}.
In Section~\ref{sec:properties}, we calculate physical properties including electron density, electron temperature, and direct-method O/H.
We derive empirical metallicity calibrations in Section~\ref{sec:results}.
Finally, in Section~\ref{sec:discussion}, we discuss these new calibrations in the context of existing literature calibrations and summarize our conclusions.

Throughout this paper, we adopt a \citet{cha2003} initial mass function (IMF), \citet{asp2021} solar abundances (12+log(O/H$)_{\odot}=8.69$), and a cosmology described by $H_0=70\mbox{ km  s}^{-1}\mbox{ Mpc}^{-1}$, $\Omega_m=0.30$, and
$\Omega_{\Lambda}=0.7$.
Emission-line wavelengths are given in the vacuum rest frame.
The term metallicity represents the gas-phase oxygen abundance unless specifically noted otherwise.
Strong-line ratios are defined and abbreviated as follows:
\begin{equation}
\text{O3} = [\text{O}~\textsc{iii}]\lambda5008/\text{H}\beta
\end{equation}
\begin{equation}
\text{O2} = [\text{O}~\textsc{ii}]\lambda3728/\text{H}\beta
\end{equation}
\begin{equation}
\text{R23} = \frac{[\text{O}~\textsc{iii}]\lambda\lambda4960,5008+[\text{O}~\textsc{ii}]\lambda3728}{\text{H}\beta}
\end{equation}
\begin{equation}
\text{O32} = [\text{O}~\textsc{iii}]\lambda5008/[\text{O}~\textsc{ii}]\lambda3728
\end{equation}
\begin{equation}
\text{Ne3O2} = [\text{Ne}~\textsc{iii}]\lambda3870/[\text{O}~\textsc{ii}]\lambda3728
\end{equation}
\begin{equation}
\text{N2} = [\text{N}~\textsc{ii}]\lambda6585/\text{H}\alpha
\end{equation}
\begin{equation}
\text{O3N2} = \frac{[\text{O}~\textsc{iii}]\lambda5008/\text{H}\beta}{[\text{N}~\textsc{ii}]\lambda6585/\text{H}\alpha}
\end{equation}
\begin{equation}
\text{N2O2} = [\text{N}~\textsc{ii}]\lambda6585/[\text{O}~\textsc{ii}]\lambda3728
\end{equation}
In this paper, [O\ii]$\lambda$3728 denotes the sum of both [O\ii]$\lambda\lambda$3727,3730 doublet components.

\section{Observations, measurements, and sample}\label{sec:obs}

\subsection{Observations and data reduction}

This analysis uses publicly-available {\it JWST}/NIRSpec Micro-Shutter Array (MSA) spectroscopic data from the CEERS survey \citep[Program ID: 1345][Finkelstein et al.,in prep.; Arrabal Haro et al., in prep.]{fin2022a,fin2022b}.
These data include observations of 6 pointings in the AEGIS field with the G140M/F100LP, G235M/F170LP, and G395M/F290LP grating/filter configurations, providing continuous wavelength coverage (excepting the chip gap) spanning $1-5$~$\mu$m at a spectral resolution of $R\sim1000$.
At each pointing, the total on-source integration in each configuration was 3107~sec.

The data were reduced to produce two-dimensional (2D) spectra from which one-dimensional (1D) flux-calibrated spectra were extracted,
 as described in \citet{sha2023} and \citet{san2023bpt}.
A wavelength-dependent slit-loss correction was then applied to the 1D spectra (see \citealt{red2023} for details).
Out of 318 total unique targets, spectroscopic redshifts were measured for 252 sources.
Eight sources were identified as active galactica nuclei (AGNs) based on the presence of broad emission lines or large [N\ii]/H$\alpha$ ratios (log(N2$)>-0.3$).
The broad-line AGNs were first identified through visual inspection of the line profiles where a narrow line
 with a broad component at the base was present, and confirmed via significant residuals at high velocity offsets when the line was fit with a
 single Gaussian profile.
At the modest spectral resolution of these observations ($R\sim1000$), clearly identifiable broad components have widths of more than
several hundred km~s$^{-1}$ that indicate the presence of an accreting supermassive black hole \citep[e.g.,][]{koc2023,mai2023}.

The remaining targets are assumed to have emission lines predominantly powered by star formation.
Stellar population properties were inferred with the spectral energy distribution (SED) fitting code Fast \citep{kri2009} by fitting the flexible stellar population synthesis models of \citet{con2009} to public multi-wavelength photometry.
We assumed a delayed-$\tau$ star-formation history and either solar stellar metallicity and the \citet{cal2000} attenuation curve or sub-solar metallicity and the SMC attenuation curve \citep{gor2003} based on the redshift and stellar mass of the source, as detailed in \citet{sha2023}.
For roughly one third of the sample for which reduced {\it JWST} NIRCam imaging was available, models were fit to combined {\it HST} and {\it JWST}/NIRCam photometry.
For the remaining two thirds, the 3D-HST survey photometric catalogs comprising {\it HST}, Spitzer, and ground-based imaging were used \citep{mom2016,ske2014}.
In both cases, the observed photometric measurements were corrected for the contribution from emission lines using the measured line fluxes from the {\it JWST}/NIRSpec spectra \citep{san2023bpt}.

\subsection{Band-to-band flux calibration}\label{sec:bandtoband}

Emission line fluxes were measured from 1D science spectra by fitting Gaussian models on top of the continuum defined by the best-fit stellar population model as described in \citet{san2023bpt}.
When fitting each emission line, the continuum model was allowed to vary by a small additive offset to fine-tune the continuum fit in a local
 wavelength region around the line.
Of the 16 CEERS targets considered here, four (IDs 1027, 792, 397, and 3788) were covered by {\it JWST}/NIRCam observations at the time of this analysis and had continuum models based on the NIRCam+HST photometry, while the remaining targets had continuum models fit to the 3D-HST photometry.
The measured line fluxes and our results do not significantly change if we instead use 3D-HST photometry for the entire sample.

This analysis requires robust emission-line ratios to calculate the reddening correction, \te, and O/H, and for accurate strong-line ratios, some of which are widely separated in wavelength (i.e., O2, R23, O32, N2O2).
While the absolute flux calibration has no impact on the results in this paper, achieving an accurate relative flux calibration is of key importance.
To ensure accurate line ratios, we first seek to use line flux measurements from within the same grating whenever possible.
If a line fell in the region of overlapping wavelength coverage between two gratings, we use the line flux measured in the same grating as the other feature(s) in a line ratio.
However, some line ratios for a subset of targets necessitate comparing line fluxes measured in different gratings depending on the redshift of the target and the wavelength separation of the lines in each ratio.
We thus took particular care with the relative flux calibration between gratings to minimize offsets between grating configurations using the following method for each target in our sample.

First, if one or more emission lines are measured in both neighboring gratings and are detected at $>5\sigma$ significance in both gratings, we use the ratio of the measured overlapping line flux(es) to place spectra in the two gratings on the same relative flux scale.
If no lines are detected in the overlap region but H$\alpha$ falls in the redder grating while H$\beta$ and higher order Balmer lines fall in the bluer grating, then we scale the redder grating such that the H$\alpha$ flux matches the expected observed (i.e., reddened) flux based on the brightness and ratios of the bluer Balmer lines.
Finally, if neither of the above cases are present, we integrate the continuum in the overlapping wavelength region between the two gratings and scale according to the ratio of the integrated fluxes.
In all cases, we scale the G140M and G395M spectra to match the G235M spectrum since the G235M configuration covers an overlapping wavelength range with both of the other two settings.
We also calculate the uncertainty on the scaling factors.
If a line ratio includes features measured in different gratings, then the uncertainty on the scaling factor is propagated into the final uncertainty on the line ratio along with the flux measurement uncertainties.
If a line ratio instead compares lines measured in the same grating, then the error on the line ratio is calculated only from the measurement uncertainty on each line flux.

It is notable that for all objects in our sample the ratios O3, Ne3O2, N2, O3N2, and [O\iii]$\lambda$4364/[O\iii]$\lambda$5008 are unaffected by band-to-band uncertainties.
All of these besides O3N2 only compare lines measured in the same grating, while for O3N2 any scaling factors cancel out since the numerator and denominator respectively include lines from a single grating.
We further determined the reddening correction (see Sec.~\ref{sec:dust} below) using only the subset of H Balmer recombination lines falling in the same grating as H$\beta$.
As such, multi-grating flux calibration uncertainties have no affect on these line ratios or our derived \te\ and 12+log(O$^{2+}$/H) values.
The line ratios impacted by band-to-band uncertainties include O2, R23, and O32 for only 4/16 targets and N2O2 for all targets with coverage of both [N\ii] and [O\ii].
Consequently, the inferred 12+log(O$^{+}$/H) values for 4/16 objects are also affected by additional uncertainty from the band-to-band calibration.
We thus find that systematics related to the relative flux calibration between grating configurations do not significantly impact our results.

\subsection{CEERS auroral-line sample selection}

We selected objects from the full CEERS/NIRSpec sample with detections of the [O\iii]$\lambda$4364 emission line using the following process.
We first selected all non-AGN sources with a measured [O\iii]$\lambda$4364 signal-to-noise ratio of S/N$\ge$2.
We then visually inspected the 2D and 1D spectra of the resulting 58 targets to select those with robust [O\iii]$\lambda$4364 by ensuring that the line is identifiable in both 2D and 1D spectra, falls at the expected wavelength according to the redshift measured from brighter lines, falls at the same spatial position in the 2D spectra as brighter lines, is morphologically well-behaved in 2D, and is not narrower than the instrumental resolution (i.e., excluding single-pixel noise spikes).
This selection results in a sample of 16 [O\iii]4364-emitters spanning $z=2.16-8.68$ with a median redshift of 4.6 (Table~\ref{tab:properties}).
The redshift distribution of this sample is shown by the green histogram in Figure~\ref{fig:zhist}.
The observed emission-line fluxes of these galaxies are presented in Table~\ref{tab:lines} in Appendix~\ref{app:lines}.
Figure~\ref{fig:oiiia} shows the region of the 2D and 1D spectra covering H$\gamma$ and \oiiia\ for these 16 galaxies.
The \oiiia\ significance spans 2.4$\sigma$ to 6.1$\sigma$ with a median S/N of 4.2.

\begin{figure}
\centering
\includegraphics[width=\columnwidth]{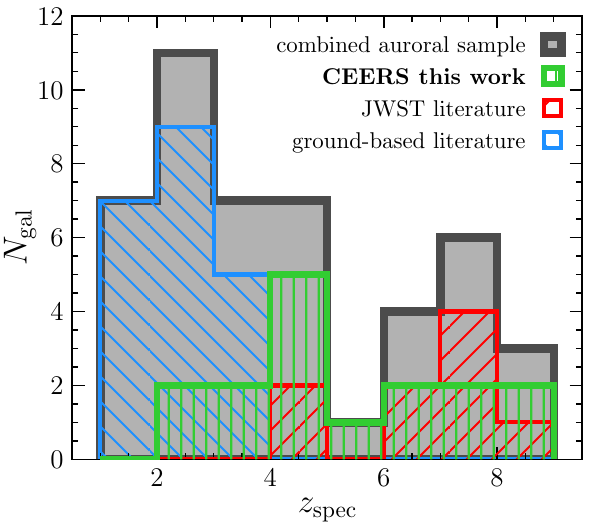}
\caption{Redshift distribution of the combined high-redshift auroral-line sample (gray) and the constituent samples from CEERS {\it JWST}/NIRSpec observations (green), additional {\it JWST}/NIRSpec auroral-line emitters in the literature (red), and sources with auroral-line detections from ground-based spectroscopy (blue).
}\label{fig:zhist}
\end{figure}

\begin{table*}
 \centering
 \caption{Properties of the CEERS auroral-line sample.
 }\label{tab:properties}
 \movetableright=-3cm
 \setlength{\tabcolsep}{2.5pt}
 \renewcommand{\arraystretch}{1.5}
 \resizebox{\textwidth}{!}{
 \begin{tabular}{ r r l l r r r r r r r }
   \hline\hline
  ID\tablenotemark{a}  &  $z_{\mathrm{spec}}$  &  R.A.  &  Dec.  &  dust lines\tablenotemark{b}  &  $E(B$$-$$V)_{\mathrm{gas}}$  &  $n_e$($[$S~\textsc{ii}$]$)  &  $T_e$($[$O~\textsc{iii}$]$)  &  12+log$\left(\frac{\mbox{O}^{2+}}{\mbox{H}^+}\right)$  &  12+log$\left(\frac{\mbox{O}^+}{\mbox{H}^+}\right)$  &  12+log(O/H)  \\
    &  &  (deg.) &  (deg.) &  &    &  (cm$^{-3}$)  &  (K)  &  &  &  \\
   \hline\hline
  1019  &  8.679  &  $215.03539$  &  $52.89066$  &  H$\beta$,H$\gamma$,H$\delta$  &  $<$$0.36$  &  ---  &  $17000$$\pm$$1790$  &  $7.75$$\pm$$0.11$  &  $6.53$$\pm$$0.12$  &  $7.78$$\pm$$0.11$  \\
  1149  &  8.175  &  $215.08971$  &  $52.96618$  &  H$\beta$,H$\gamma$,H$\delta$  &  $0.22$$\pm$$0.21$  &  ---  &  $16730$$\pm$$2910$  &  $7.75$$\pm$$0.19$  &  $6.95$$\pm$$0.19$  &  $7.82$$\pm$$0.18$  \\
  1027  &  7.819  &  $214.88299$  &  $52.84042$  &  H$\beta$,H$\gamma$  &  $0.22$$\pm$$0.21$  &  ---  &  $18970$$\pm$$2300$  &  $7.60$$\pm$$0.13$  &  $6.20$$\pm$$0.17$  &  $7.62$$\pm$$0.13$  \\
  698  &  7.470  &  $215.05032$  &  $53.00744$  &  H$\beta$,H$\gamma$,H$\delta$  &  $0.04$$\pm$$0.20$  &  ---  &  $13330$$\pm$$2420$  &  $8.01$$\pm$$0.23$  &  $6.86$$\pm$$0.23$  &  $8.04$$\pm$$0.23$  \\
  792  &  6.257  &  $214.87177$  &  $52.83317$  &  H$\alpha$,H$\beta$,H$\gamma$,H$\delta$  &  $<$$0.41$  &  ---  &  $28310$$\pm$$3380$  &  $7.41$$\pm$$0.11$  &  $6.64$$\pm$$0.21$  &  $7.48$$\pm$$0.12$  \\
  397  &  6.000  &  $214.83620$  &  $52.88269$  &  H$\alpha$,H$\beta$,H$\gamma$  &  $0.06$$\pm$$0.02$  &  ---  &  $14230$$\pm$$1240$  &  $7.92$$\pm$$0.10$  &  $7.03$$\pm$$0.12$  &  $7.98$$\pm$$0.10$  \\
  1536  &  5.033  &  $214.97723$  &  $52.94078$  &  H$\beta$,H$\gamma$,H$\delta$  &  $<$$0.41$  &  ---  &  $24100$$\pm$$4320$  &  $7.45$$\pm$$0.14$  &  $6.40$$\pm$$0.18$  &  $7.48$$\pm$$0.15$  \\
  1477  &  4.631  &  $215.00349$  &  $52.96954$  &  H$\beta$,H$\gamma$,H$\delta$  &  $0.08$$\pm$$0.16$  &  ---  &  $18020$$\pm$$2220$  &  $7.64$$\pm$$0.13$  &  $6.92$$\pm$$0.13$  &  $7.72$$\pm$$0.12$  \\
  1746  &  4.560  &  $215.05401$  &  $52.95687$  &  H$\beta$,H$\gamma$,H$\delta$  &  $<$$0.18$  &  $1020$$\pm$$1590$  &  $15570$$\pm$$1970$  &  $7.88$$\pm$$0.14$  &  $7.14$$\pm$$0.14$  &  $7.95$$\pm$$0.14$  \\
  1665  &  4.482  &  $215.17820$  &  $53.05935$  &  H$\beta$,H$\gamma$,H$\delta$  &  $0.17$$\pm$$0.12$  &  $300$$\pm$$400$  &  $11730$$\pm$$1310$  &  $8.16$$\pm$$0.15$  &  $7.63$$\pm$$0.14$  &  $8.27$$\pm$$0.15$  \\
  1559  &  4.471  &  $215.06486$  &  $52.99984$  &  H$\beta$,H$\gamma$,H$\delta$  &  $<$$0.21$  &  ---  &  $18130$$\pm$$2950$  &  $7.86$$\pm$$0.16$  &  $6.60$$\pm$$0.20$  &  $7.88$$\pm$$0.16$  \\
  1651  &  4.375  &  $215.16922$  &  $53.05477$  &  H$\beta$,H$\gamma$  &  $<$$0.61$  &  ---  &  $16220$$\pm$$3270$  &  $7.68$$\pm$$0.21$  &  $7.03$$\pm$$0.24$  &  $7.77$$\pm$$0.21$  \\
  11728  &  3.869  &  $215.08487$  &  $52.97074$  &  H$\beta$,H$\gamma$,H$\delta$  &  $0.11$$\pm$$0.12$  &  ---  &  $20370$$\pm$$2580$  &  $7.44$$\pm$$0.12$  &  $6.23$$\pm$$0.17$  &  $7.46$$\pm$$0.12$  \\
  11088  &  3.302  &  $214.93421$  &  $52.82637$  &  H$\alpha$,H$\beta$,H$\gamma$,H$\delta$  &  $0.34$$\pm$$0.01$  &  $240$$\pm$$260$  &  $12160$$\pm$$1630$  &  $8.05$$\pm$$0.18$  &  $8.01$$\pm$$0.18$  &  $8.33$$\pm$$0.14$  \\
  3788  &  2.295  &  $214.89079$  &  $52.86870$  &  H$\beta$,H$\gamma$,H$\delta$  &  $0.06$$\pm$$0.09$  &  $<$$1130$  &  $12130$$\pm$$970$  &  $8.12$$\pm$$0.10$  &  $7.56$$\pm$$0.41$  &  $8.23$$\pm$$0.14$  \\
  3537  &  2.162  &  $215.14363$  &  $53.00480$  &  H$\beta$,H$\gamma$,H$\delta$  &  $0.26$$\pm$$0.08$  &  ---  &  $24510$$\pm$$3430$  &  $7.10$$\pm$$0.11$  &  $5.72$$\pm$$0.18$  &  $7.12$$\pm$$0.11$  \\
   \hline\hline
 \end{tabular}}
 \tablenotetext{a}{CEERS ID number.}
 \tablenotetext{b}{Balmer lines used to calculate \ebvgas.}
\end{table*}

\begin{figure*}
\centering
\includegraphics[width=\textwidth]{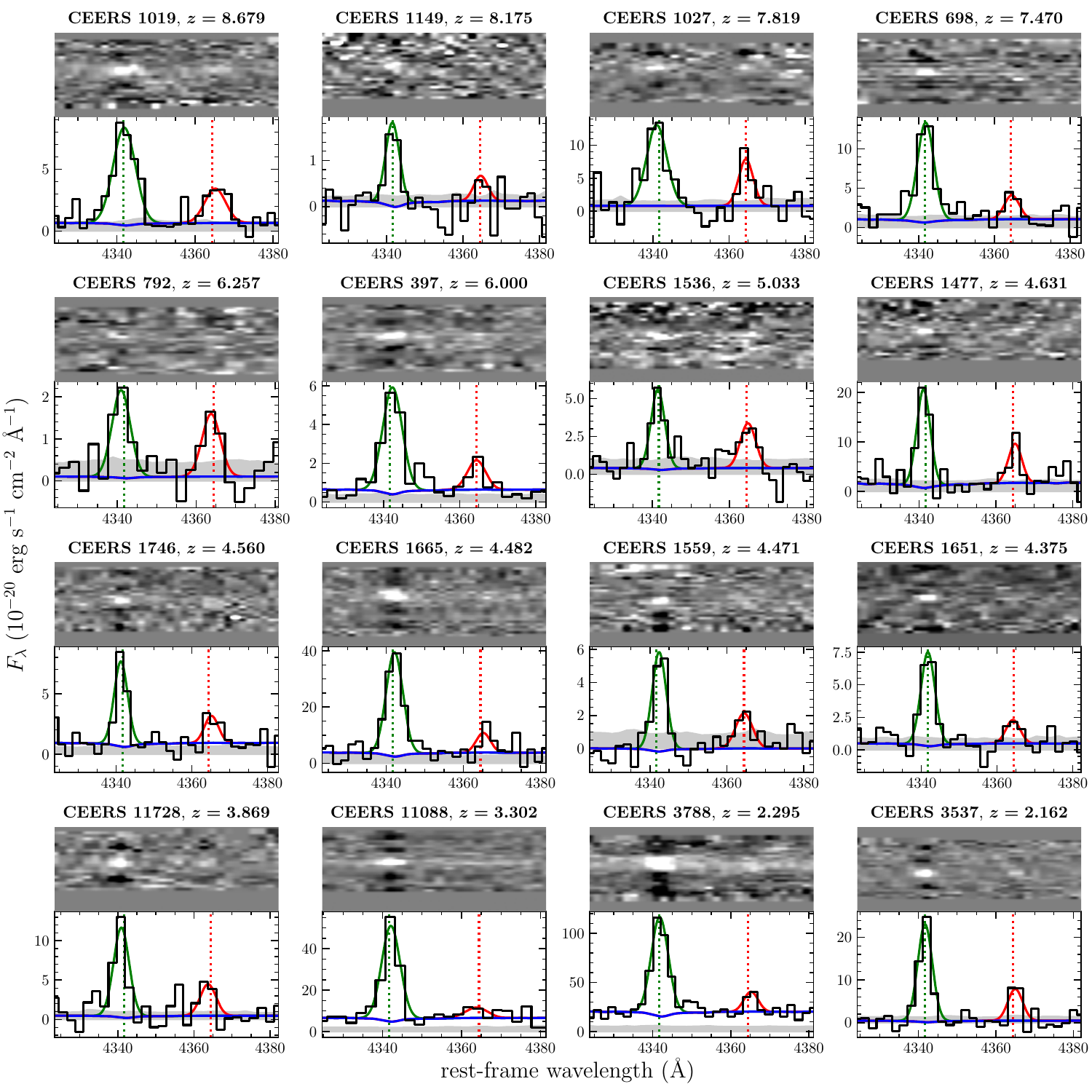}
\caption{1D and 2D spectra displaying the detected [O\iii]$\lambda$4364 emission lines and H$\gamma$ for the 16 galaxies in the CEERS auroral-line sample.
The black line displays the 1D science spectrum, while the gray shaded region shows the 1D error spectrum.
The blue, green, and red solid lines display the best-fit continuum, and H$\gamma$ and [O\iii]$\lambda$4364 line profiles, respecitvely.
The dotted vertical lines show the rest-frame wavelengths of these transitions.
}\label{fig:oiiia}
\end{figure*}

\begin{figure}
\centering
\includegraphics[width=\columnwidth]{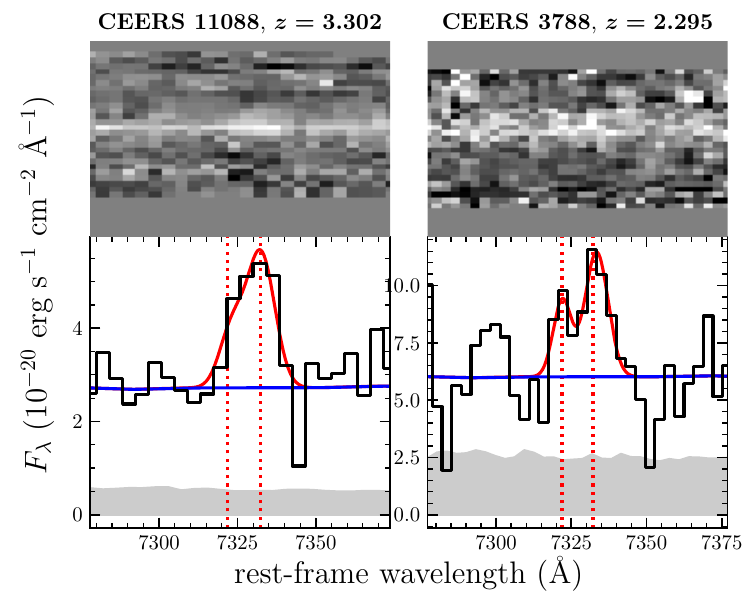}
\caption{1D and 2D spectra showing detections of \oiia\ for two CEERS targets.
The blue line shows the best-fit continuum model, while the red line shows the combined fit to [O\ii]$\lambda7322$ and [O\ii]$\lambda$7332.
}\label{fig:oiia}
\end{figure}

Two of the CEERS \oiiia-emitters, 11088 ($z=3.302$) and 3788 ($z=2.295$), also display detections of the auroral [O\ii]$\lambda\lambda$7322,7332 emission line doublet, shown in Figure~\ref{fig:oiia}.
The first detections of the [O\ii] auroral lines at high redshift were recently reported by \citet{san2023oiia}.
To our knowledge, these new detections represent only the second time \oiia\ has been reported beyond the low-redshift universe.
We will use these [O\ii] auroral lines to constrain \te\ in the low-ionization nebular zone for these two objects.

\subsection{Literature {\it JWST} auroral-line sample}

To supplement the CEERS auroral sample, we selected 9 additional targets from the literature with \te-based metallicities and detected auroral emission lines from {\it JWST} spectroscopy.
Four of these targets have $R\sim1000$ NIRSpec data from the ERO program targeting the SMACS~0723 cluster.
We use the published line fluxes from \citet{cur2023} for ERO IDs 4590, 6355, and 10612, and line measurements from \citet{nak2023} for ERO ID 5144.\footnote{If we instead use the line fluxes reported by \citet{nak2023} for ERO IDs 4590, 6355, and 10612, the derived 12+log(O/H) values differ by only $0.05-0.13$~dex representing changes of $\lesssim1\sigma$ from our fiducial values.}
Five additional sources have published auroral line detections measured from $R\sim2700$ NIRSpec observations from the GLASS ERS program \citep{tre2022}.
We utilize line measurements from \citet{nak2023} for GLASS IDs 100003, 10021, 150029, and 160133, and the line fluxes from \citet{jon2023} for GLASS ID 150008.
The {\it JWST} literature auroral-line sample has redshifts spanning $z=4.01-8.50$ with a median redshift of 7.29, the distribution of which is displayed by the red histogram in Figure~\ref{fig:zhist}.
The detected auroral line is O\iii]$\lambda$1666 for GLASS 150008 and [O\iii]4364 for the 8 other {\it JWST} literature galaxies.
These literature auroral lines have S/N$=3.3-9.6$ with a median significance of 6.0.

\subsection{Ground-based auroral-line sample}

We also include a sample of galaxies at $z>1$ with auroral-line detections from ground-based spectroscopy.
This sample is predominantly made up of the sample presented in \citet{san2020} that totals 18 targets including several from literature sources \citep{vil2004,bra2012b,chr2012a,chr2012b,sta2013,sta2014,bay2014,jam2014,san2016b,koj2017,ber2018}.
We supplement this sample with a \oiiia-detected galaxy from \citet{gbu2019} and two galaxies with detected [O\ii] auroral emission lines presented in \citet{san2023oiia}.
We do not include composite spectra \citep[e.g.,][]{ste2016,gbu2022}, but instead limit to individual sources.
The ground-based auroral-line sample thus includes 21 galaxies with a redshift distribution shown by the blue histogram in Figure~\ref{fig:zhist}, spanning $z=1.42-3.63$.
This ground-based sample includes 8 galaxies with \oiiia\ detections, 11 with O\iii]$\lambda$1666 detections, and 2 with [O\ii]$\lambda\lambda$7322,32 detections.
For the ground-based sample, we adopt the reddening-corrected line ratios, \te, and direct-method metallicities calculated by \citet{san2020} and \citet{san2023oiia} that were derived using a methodology consistent with this work.

\subsection{Combined high-redshift auroral-line sample}

To obtain sufficient statistics to construct new empirical high-redshift metallicity calibrations, we combine the CEERS, {\it JWST} literature, and ground-based literature auroral-line samples, resulting in a combined high-redshift auroral-line sample consisting of 46 unique sources with \te-based metallicities.
The gray histogram in Figure~\ref{fig:zhist} shows the redshift distribution of the combined sample spanning $z=1.4-8.7$, which has a median redshift of $z_{\mathrm{med}}=3.63$.
The currently available data do not suggest strong evolution of ISM ionization conditions or metallicity calibrations over $z\sim2-9$, implying that galaxies over this large redshift range may be used in a single calibrating sample \citep[see Sec.~\ref{sec:zrange};][]{san2023bpt}.
Of these 46 galaxies, all have detected O3, 39 have detected O2, 43 have detected R23, 39 have detected O32, 31 have detected Ne3O2, and 12 have detected N2, O3N2, and N2O2.
This sample is more than twice the size of the largest high-redshift auroral-line compilation assembled to-date \citep{san2023oiia}.
The combined sample has a median S/N of 4.8 for the auroral line(s) used to derive \te.
While 6 targets have auroral lines that are only marginally detected at $2-3\sigma$, excluding these targets from the analysis did not significanlty change the best-fit calibrations derived from this sample.

\section{Derived physical properties}\label{sec:properties}

In this section, we describe how physical properties including dust reddening, emission-line ratios, electron density and temperature, and oxygen abundances were calculated for the CEERS and literature {\it JWST}/NIRSpec auroral-line samples.

\subsection{Dust reddening, SFR, and emission-line ratios}\label{sec:dust}

A robust correction for dust reddening is required for accurate \te\ and O/H inferences.
We derived the nebular reddening, \ebvgas, using the observed ratios of H Balmer recombination lines including H$\alpha$, H$\beta$, H$\gamma$, and H$\delta$ assuming the Milky Way (MW) extinction law of \citet{car1989}.
The nebular attenuation curve derived directly for $z\sim2$ star-forming galaxies is consistent with the MW curve \citep{red2020}.
Furthermore, analysis of Balmer and Paschen line ratios from {\it JWST} at $z=1-3$ do not indicate significant deviation from the MW law \citep{red2023}.
To reduce uncertainty due to the relative flux calibration between gratings (Sec.~\ref{sec:bandtoband}), we only used Balmer line fluxes measured in the same grating as H$\beta$ to derive the nebular reddening.
The set of lines employed for each of the CEERS auroral-line emitters is reported in Table~\ref{tab:properties}.
For the {\it JWST} literature sources, we do not have complete information about whether the reported Balmer lines were measured in different spectroscopic configurations and simply used the subset of H$\alpha$, H$\beta$, H$\gamma$, and H$\delta$ that are detected at $\ge3\sigma$.

\ebvgas\ was calculated via a $\chi^2$ minimization routine that simultaneously fits to the set of available ratios out of H$\alpha$/H$\beta$, H$\gamma$/H$\beta$, and H$\delta$/H$\beta$, taking into account the uncertainty on each observed ratio.
The intrinsic ratios were calculated with \texttt{pyneb} \citep{lur2015} assuming \te=15,000~K, a typical value for our sample.
The derived \ebvgas\ values, reported in Table~\ref{tab:properties}, were then used to dust-correct the observed line fluxes based on their rest-frame wavelengths assuming the \citet{car1989} extinction law.
The sample is generally not significantly dusty, with \ebvgas$<0.3$ for the vast majority of targets, such that our results are not strongly dependent on the reddening correction.

\begin{figure}
\centering
\includegraphics[width=\columnwidth]{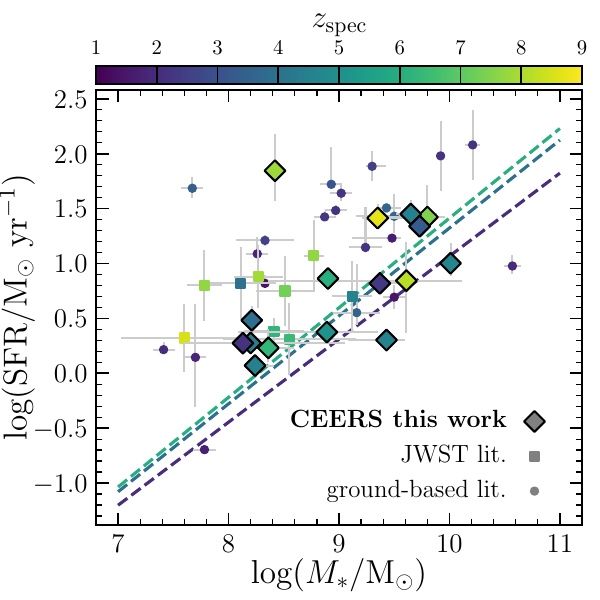}
\caption{SFR vs.\ \mstar\ for the CEERS, {\it JWST} literature, and ground-based auroral-line samples, color-coded by redshift.
The dashed lines show the mean star-forming main sequence parameterization of \citet{spe2014} evaluated at $z=2$, $z=4$, and $z=6$ on the same color scale.
All SFRs are derived from dust-corrected H$\alpha$ or H$\beta$ luminosity.
Literature data and the \citet{spe2014} relation have been converted to a \citet{cha2003} IMF, and their SFRs have been lowered by 0.34~dex to account for the difference between the solar-metallicity conversion factors used in those works and the low-metallicity BPASS binary conversion factor employed here \citep{sha2023}.
}\label{fig:sfrmstar}
\end{figure}

SFRs were calculated using dust-corrected H$\alpha$ luminosity when H$\alpha$ was covered, otherwise dust-corrected H$\beta$ luminosity was employed.
We adopt the conversion factor from Balmer line luminosity to SFR based on $Z_*=0.001$ BPASS binary stellar population synthesis models \citep{eld2017} appropriate for moderate and low metallicity high-redshift systems \citep{red2022,sha2023}.
SFR as a function of \mstar\ is shown in Figure~\ref{fig:sfrmstar} for the CEERS, {\it JWST} literature, and ground-based auroral-line samples, color-coded by redshift.
The vast majority of the auroral-line detected high-redshift galaxies lie above the mean star-forming main sequence at their respective epoch \citep{spe2014}.
This bias toward high specific SFR (sSFR=SFR/\mstar) is a result of selecting sources based on detections of weak emission lines.
Such an effect may systematically bias derived metallicity calibrations since the ionization conditions within the ISM may differ
 between galaxies on and significantly above the main sequence, though precisely how remains an open question at high redshifts
 (see Sec.~\ref{sec:areas} for further discussion).

Emission-line ratios were calculated using the reddening-corrected line fluxes.
Due to the close proximity of the involved lines, the ratios O3, Ne3O2, N2, and O3N2 have virtually no dependence on the reddening correction.
In contrast, the final R23, O2, O32, N2O2, [O\iii]$\lambda$4364/[O\iii]$\lambda$5008, and \oiia/[O\ii]$\lambda$3728 ratios depend on the inferred \ebvgas.
We define a detection of a strong-line ratio as the case where all lines in that ratio are detected at $\ge3\sigma$ significance.
In the case that one or more lines in a ratio do not meet this criterion, we calculate 3$\sigma$ limits when possible.

We adopt the following two exceptions to these criteria for detections of line ratios.
First, since the numerator of R23 is the sum of
 [O\ii]$\lambda$3728 and [O\iii]$\lambda\lambda$4960,5008, R23 is also well constrained in the case where 
 [O\iii]$\lambda\lambda$4960,5008 and H$\beta$ are detected while
 [O\ii]$\lambda$3728 is not detected (S/N$<$3) but the upper limit on [O\ii] flux is significantly less than the
 [O\iii]$\lambda\lambda$4960,5008 flux.
Accordingly, we also consider the R23 ratio to be detected when [O\ii]$\lambda$3728 is not detected but the 3$\sigma$ upper limit on
 its flux is less than 5 times the detected [O\iii]$\lambda$5008 flux.
In this scenario, $\mbox{R}23\approx[\mbox{O~\textsc{iii}}]\lambda\lambda4960,5008/H\beta$ is a good approximation, and the uncertainty
 on R23 resulting from neglecting [O\ii] is $<0.06$~dex and is included in the final calculated uncertainty.
This level of uncertainty is acceptable given that 9 targets with detected [O\ii] have $\sigma(R23)\ge0.05$~dex.
Four targets with undetected [O\ii] are included in the R23-detected sample based on this selection, in addition to 39 targets for
 which all of the necessary lines for R23 are detected at S/N$\ge$3.

Second, since the sample contains very few galaxies at extremely low metallicities (12+log(O/H$)<7.5$), we relax the detection threshold
 for [O\ii]$\lambda$3728 that is typically weak in such metal-poor galaxies, considering [O\ii] as detected when S/N$\ge$2 in this regime.
This relaxed criterion allows for the inclusion of the lowest metallicity galaxy in our sample, ERO~ID~4590, for line ratios involving [O\ii].
However, our best-fit calibrations for O2, R23, O32, and Ne3O2 do not significantly change if ERO~ID~4590 is instead excluded. 

The vast majority of the combined high-redshift sample ($\ge$39/46) is detected in the O3, O2, R23, and O32 ratios, while Ne3O2 is detected for 31/46 sources.
As such, the detected sample for line ratios based on $\alpha$-element metals (i.e., O and Ne) is reasonably representative of the full combined auroral-line sample.

[N\ii]$\lambda$6585 is covered in the spectra of 11/25 {\it JWST} targets.
With {\it JWST}/NIRSpec, H$\alpha$ and [N\ii] can only be accessed out to $z\approx6.7$.
Nine of the {\it JWST} CEERS and literature sources are at $z>6.7$ such that [N\ii] was not covered, while for 5 other objects [N\ii] fell in the chip gap (CEERS 1651), fell off the detector due to the position on the MSA mask, or was not reported in the literature reference.
Of the {\it JWST} targets with coverage, [N\ii] is detected for 6 galaxies.
This low detection rate is likely due to the low [N\ii]/H$\alpha$ ratios that appear to be typical of metal-poor high-redshift galaxies.
\citet{san2023bpt} and \citet{sha2023b} used composite spectra of CEERS targets to demonstrate that, at $z>2.7$, it is common for [N\ii] to be $10-30$ times weaker than H$\alpha$, and indeed sometimes weaker even than \oiiia\ as demonstrated by this sample.
[N\ii] statistics are similarly poor in the ground-based sample, with 10/21 sources lacking [N\ii] coverage and 6 [N\ii] detections in total \citep{san2020,san2023oiia}.
Consequently, only 12/46 sources in the combined sample have detections of the N2, O3N2, and N2O2 ratios.

\subsection{Electron density}\label{sec:den}

The electron density, \den, is derived from the ratio of the components of the [S\ii]$\lambda\lambda$6718,6733 doublet when both lines are detected.
The spectral resolution of $R\sim1000$ offered by the medium-resolution NIRSpec gratings is insufficient to resolve the components of the [O\ii]$\lambda\lambda$3727,3730 doublet such that \den\ cannot be reliably constrained using [O\ii] in the CEERS or ERO SMACS 0723 data, though observations taken with the $R\sim2700$ high-resolution NIRSpec gratings used in GLASS are sufficient.
If the spectral resolution is too low to Nyquist sample the separation between the [O\ii] doublet members, the inferred doublet ratio is biased toward unity (\citealt{san2016}).
The \texttt{pyneb} Python package was used to calculate \den([S\ii]) using the S$^+$ collision strengths of \citet{tay2010} assuming \te=15,000~K, though this calculation is very weakly dependent on \te.

Both components of the [S\ii] doublet were detected with S/N$\ge$3 for 4 objects in the CEERS auroral-line sample, with the inferred densities ranging from the low-density limit to $n_{\mathrm{e}}\approx1000$~cm$^{-3}$ with large uncertainties (Table~\ref{tab:properties}).
\citet{iso2023} report \den([O\ii]) for GLASS 150029 and 160133, finding values of 158~cm$^{-3}$ and 234~cm$^{-3}$, respectively, yielding a total of 6/25 {\it JWST} sources in our samples with \den\ estimates.
The same authors also report [O\ii] densities for several of the CEERS auroral-line targets, but these \den\ constraints are unreliable because the separate components of the [O\ii] doublet are not resolved at $R\sim1000$.
In the ground-based auroral-line sample, 17/21 sources have \den\ measurements based on [O\ii] from $R\ge3000$ spectra or [S\ii], for which \den\ ranges from the low-density limit to 2900~cm$^{-3}$ with a median value of 280~cm$^{-3}$ \citep{san2020,san2023oiia}.
If the 6 {\it JWST} sources with \den\ constraints are included, the median density for 23/46 objects in the combined sample is 278~cm$^{-3}$.
This value is in good agreement with the typical electron density found for large samples of $z\sim2-3$ star-forming galaxies of $250-300$~cm$^{-3}$ \citep{san2016,str2017}.
Accordingly, we assume \den=300~cm$^{-3}$ for the \te\ and abundance ratio calculations described below.
The exact assumed value has negligible impact on the metallicity results since \den\ variation changes derived \te\ values by $\lesssim1$\% when \den$<$3000~cm$^{-3}$ \citep{san2020}.

\subsection{Electron temperature}\label{sec:tem}

For all but one {\it JWST} target, the electron temperature of the high-ionization O$^{2+}$ zone of the nebula, \te([O\iii]), was calculated from the \oiiia/\oiiis\ ratio.
We used \texttt{pyneb} with the O$^{2+}$ collision strengths from \citet{sto2014}.
For GLASS 150008, \oiiia\ fell in the chip gap, but O\iii]$\lambda$1666 was significantly detected \citep{jon2023}.
We use the O\iii]$\lambda$1666/\oiiis\ ratio to calculate \te\ for this galaxy.
We use the \citet{agg1999} collision strengths for the O\iii]$\lambda$1666 calculation since it requires a 6 level atom while \citet{sto2014} only include 5 levels, following the approach used in \citet{san2020}.
The derived \te\ and O/H values change by much less than 1$\sigma$ if we instead use \citet{agg1999} for the entire sample.
In the CEERS sample, \te([O\iii]) ranges from 11,000~K to 28,000~K, as reported in Table~\ref{tab:properties}.
Four CEERS targets have very high \te\ in excess of 20,000~K (though with large uncertainties of $\sim3,500$~K),
 as hot as extremely metal-poor ($<0.1~Z_{\odot}$) local galaxies \citep[e.g.,][]{izo2012,izo2018} but similar to what has been inferred from early {\it JWST} spectra at $z\sim6-8$ \citep[e.g.,][]{cur2023,sch2022,are2022}.

The electron temperature in the low-ionization O$^+$ zone, \te([O\ii]), is required to compute O$^+$/H.
Two objects in the CEERS sample (11088 and 3788) have detections of \oiia\ (Fig.~\ref{fig:oiia}), for which we derive \te([O\ii]) with \texttt{pyneb} using the O$^+$ collision strengths of \citet{kis2009}.
These targets represent the first high-redshift sources with direct constraints on \te\ in both the low- and high-ionization zones.
We find \te([O\ii]) of $9710^{+880}_{-850}$~K for 11088 and $12,140^{+2940}_{-2790}$~K for 3788.
These values are generally similar to their high-ionization temperatures of \te([O\iii]$)\approx12,000$~K.

For the vast majority of the {\it JWST} sample, a low-ionization auroral line (e.g., \oiia) is not detected such that \te([O\ii]) cannot be directly calculated.
Following the common practice in local-universe abundance studies, we adopt a parameterized relation of \te([O\ii]) as a function of \te([O\iii]).
We use the relation of \citet{cam1986}:
\begin{equation}\label{eq:t2t3}
T_{\mathrm{e}}(\mathrm{O}~\textsc{ii}) = 0.7\times T_{\mathrm{e}}(\mathrm{O}~\textsc{iii}) + 3,000~\mathrm{K}
\end{equation}
We use this relation to infer \te([O\ii]) for all objects in the sample with only direct \te([O\iii]) measurements.
Of the two objects with measured \te([O\ii]) and \te([O\iii]), 11088 is offset $1.3\sigma$ from this line, while 3788 is consistent at the $<1\sigma$ level, suggesting that this relation is reasonable.
However, there is notable uncertainty about the form of the \te([O\ii])$-$\te([O\iii]) relation even at $z=0$, and the relation appears to have a large intrinsic scatter \citep{rog2021}.
Our results do not significantly change if we instead assume \te([O\ii])=\te([O\iii]).

In the ground-based literature sources, \te([O\iii]) is based on \oiiia\ for 8 objects and O\iii]$\lambda$1666 for 11 objects.
The remaining two ground-based objects have \te([O\ii]) measurements from \oiia.
The same \te([O\ii])$-$\te([O\iii]) relation was adopted in the ground-based literature analyses.

\subsection{Ionic and total oxygen abundances}\label{sec:oh}

Ionic and total oxygen abundances were calculated using \texttt{pyneb} with the collision strengths of \citet{sto2014} for O$^{2+}$ and \citet{kis2009} for O$^+$.
We assume that all O is in either the O$^{2+}$ or O$^+$ states inside H\ii\ regions, such that
\begin{equation}
\frac{\mathrm{O}}{\mathrm{H}} = \frac{\mathrm{O}^{2+}}{\mathrm{H}^+} + \frac{\mathrm{O}^+}{\mathrm{H}^+}
\end{equation}
The O$^{3+}$ state makes up $\lesssim$5\% of the total O even in extremely high-ionization systems \citep{ber2018,ber2021} and is thus negligible relative to the typical uncertainties on O/H in this study.
The O$^{2+}$/H$^+$ ratio is inferred from \oiiis/H$\beta$ using \te([O\iii]), while O$^+$/H$^+$ is derived from the dust-corrected [O\ii]$\lambda$3728/H$\beta$ ratio assuming the directly-constrained \te([O\ii]) when available or \te([O\ii]) calculated with equation~\ref{eq:t2t3} otherwise.
Two {\it JWST} targets lack [O\ii]$\lambda$3728 coverage: CEERS 1651 and GLASS 150008.
For CEERS 1651, [O\ii] falls in the chip gap in the G235M observations, while the position of GLASS 150008 on the NIRSpec MSA mask was such that [O\ii] fell off the detector \citep{jon2023}.
For the metallicity calculations of these two targets, we infer the dust-corrected [O\ii]$\lambda$3728 flux from the dust-corrected \oiiis\ flux assuming O32=5, a typical value for the {\it JWST} auroral-line sources, while O32 values uniformly distributed between 2 and 10 were adopted in the uncertainty calculations.
The direct-method oxygen abundances are reported in Table~\ref{tab:properties} for the CEERS auroral-line sample.
Metallicities of the CEERS objects range from 12+log(O/H$)=7.1-8.3$ with a median value of 12+log(O/H$)=7.8$ (0.13~$Z_{\odot}$), indicating that the high-redshift galaxies in this sample are relatively metal-poor.


\subsection{Uncertainties on derived properties}

Uncertainties on \ebvgas, emission-line ratios, \den, \te, and abundance ratios were calculated by perturbing the observed line fluxes according to the measured uncertainties and recalculating all of the properties based on the new realization of line strengths.
This process was repeated 500 times to sample the distribution of each property, and the 1$\sigma$ error was inferred from the 68th-percentile bounds on each quantity.
Uncertainties of line ratios comparing lines measured in different NIRSpec gratings additionally include the uncertainty on the relative flux calibration between gratings (Sec.~\ref{sec:bandtoband}).

\section{Empirical high-redshift metallicity calibrations}\label{sec:results}

\begin{figure*}
\centering
\includegraphics[width=\textwidth]{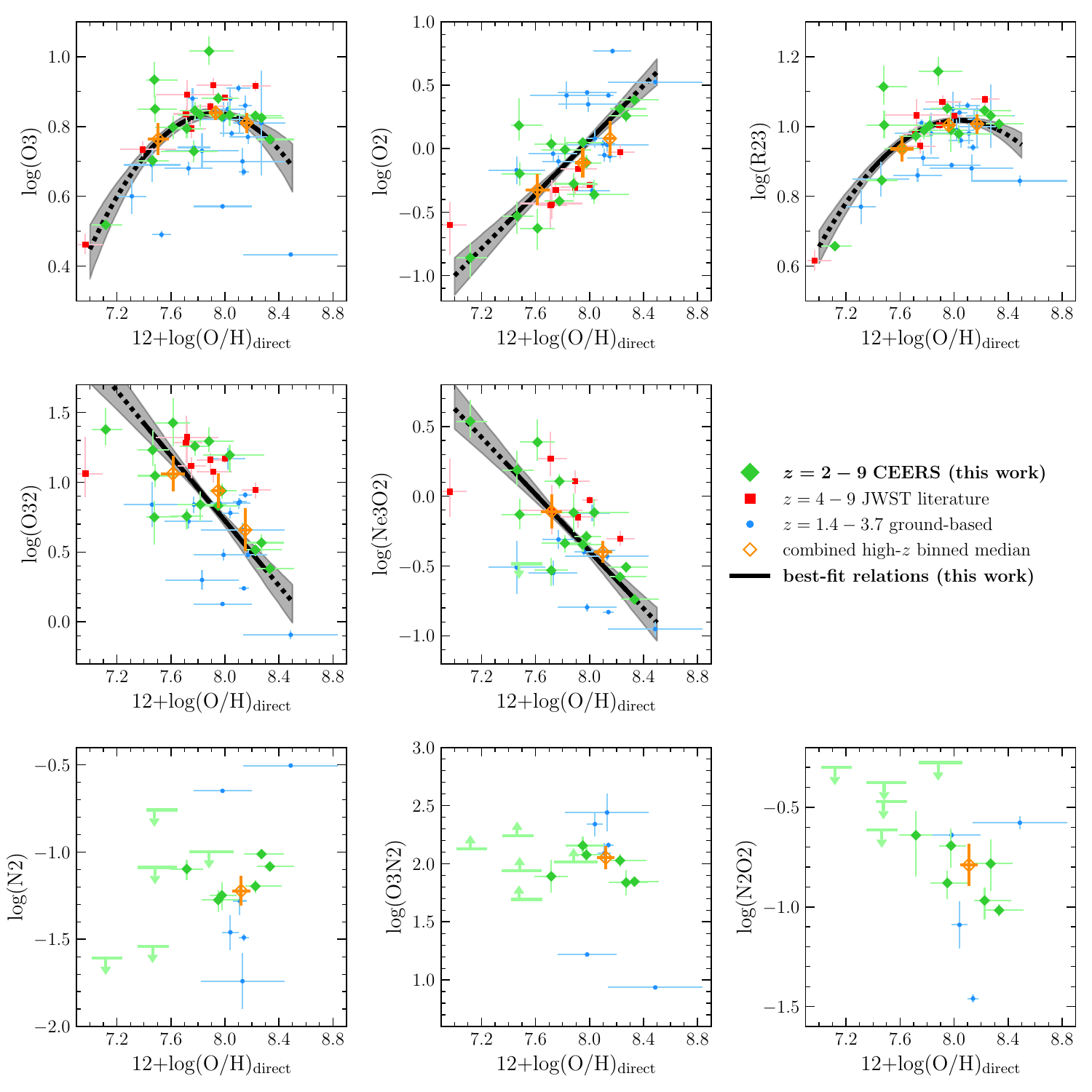}
\caption{Strong-line ratios vs.\ direct-method oxygen abundance. The CEERS, {\it JWST} literature, and ground-based \te\ samples are shown in green, red, and blue, respectively.
Orange diamonds represent median values in bins of O/H for the combined high-redshift sample.
The black lines in the top two rows display the best-fit calibrations (Table~\ref{tab:cal} and equation~\ref{eq:cal}),
 where the solid line represents the metallicity range over which the calibrations are valid with better than 0.1~dex precision ($7.4<12+\mbox{log(O/H)}<8.3$)
 while the dotted line shows the range where the calibrations are more uncertain due to small number statistics.
The gray shaded region shows the $1\sigma$ confidence interval on the best-fit relations.
}\label{fig:cals}
\end{figure*}

We now use the direct-method metallicities for the combined high-redshift sample of 46 galaxies at $z=1.4-8.7$ to construct the first empirical \te-based metallicity calibrations derived directly from high-redshift sources.
Figure~\ref{fig:cals} shows the strong-line ratios O3, O2, R23, O32, Ne3O2, N2, O3N2, and N2O2 as a function of direct-method O/H.
O3 and R23 are known to be double-valued in $z=0$ samples and local analogs, with a turnover point at roughly 12+log(O/H$)\sim8.0$ \citep[e.g.,][]{cur2020,bia2018}.
The high-redshift data are consistent with such a shape, in particular showing a dropoff toward lower O3 and R23 with decreasing metallicity at 12+log(O/H$)<7.6$.
We do not observe any signs of flattening in the O2, O3, and Ne3O2 vs.\ O/H diagrams.
A Spearman correlation test on the detected sources in each of these panels indicates the presence of significant correlations, with a correlation coefficient of $\rho_{\mathrm{s}}=0.625$ and a $p$-value of $2.1\times10^{-5}$ for O2, $\rho_{\mathrm{s}}=-0.497$ and $p$-value=$1.3\times10^{-3}$ for O32, and $\rho_{\mathrm{s}}=-0.578$ and $p$-value=$6.6\times10^{-4}$ for Ne3O2.
As discussed above, [N\ii] measurements were not available for most of the sample and only 12 sources have [N\ii] detections, such that the statistics are poor for N2, O3N2, and N2O2.
With the limited data available, no clear trends in the N-based line ratios are apparent as a function of O/H.

To further discern how these strong-line ratios depend on metallicity, we calculated medians in bins of O/H including only the galaxies with line ratio detections in each panel.
We aim to have $12-15$ galaxies per bin to obtain robust sample-averages.
Accordingly, we use 3 bins for O3, O2, R23, and O32; 2 bins for Ne3O2, and 1 bin for the N-based line ratios.
The binned medians make the turnover of O3 and R23 clear, while suggesting monotonic metallicity dependence for O32 and Ne3O2.

We fit strong-line ratio as a function of metallicity, adopting polynomial functional forms of different orders, represented as
\begin{equation}\label{eq:cal}
\log(R) = \sum_i c_i \times x^i
\end{equation}
where $x=12+\log(\mathrm{O/H})-8.0=Z/5Z_{\odot}$, $R$ is the strong-line ratio, and the coefficients $c_i$ are determined from fitting.
For each line ratio, fits are carried out on the subsample that is detected in that ratio.
We adopt a second-order polynomial for O3 and R23, and a first-order (i.e., linear) relation for O2, O32, and Ne3O2, motivated by the trends in the binned medians as well as the shape of calibrations based on these line ratios in the local universe \citep[e.g.,][]{cur2020,bia2018}.
The low number of detections and lack of a clear trend precludes fitting calibrations for N2, O3N2, and N2O2.

The best-fit calibrations are derived as follows.
For each line ratio, we fit the individually-detected sources using an orthogonal distance regression (ODR).
While ODR fitting can include inverse-variance weighting in both variables, we do not weight according to the uncertainties while fitting.
Strong-line calibrations are known to have large intrinsic scatter due to the variation of physical properties such as ionization parameter at fixed metallicity \citep[e.g.,][]{pil2016}.
This scatter is typically $0.1-0.3$~dex in line ratio at fixed O/H in local calibration samples, larger than the measurement uncertainty for most of the high-redshift sample.
As such, any targets with very small error bars should not be more heavily weighted as the intrinsic physical scatter is still large, otherwise the outcome will be biased.
Preventing this bias is especially important since the uncertainties on $R$ and O/H vary widely across our sample.
However, the uncertainties should still affect the error on the final best-fit coefficients.
To achieve this goal while preventing an unrealistic over-weighting of objects with very high S/N, we perturb the data points according to their uncertainties 500 times and fit each of the realizations.
Among the 500 realizations of best-fit relations, we compute the median $R$ as a function of O/H of the functional fits and then infer the final best-fit coefficients by fitting the functional form to the resulting curve.

The best-fit calibrations for O3, O2, R23, O32, and Ne3O2 are shown by the black lines in Figure~\ref{fig:cals}, and the best-fit coefficients are reported in Table~\ref{tab:cal}.
The best-fit relations derived in this way show good agreement with the binned medians.
In contrast, fitting with inverse-variance weighting failed to match the median trends due to the effect described above.
The gray shaded regions show the $1\sigma$ confidence interval on the best-fit calibrations, derived from the 500 realizations.
The typical uncertainty of these calibrations due to measurement uncertainties, parameterized as the uncertainty in $R$ at fixed O/H, is $\approx0.05$~dex.

We quantify the metallicity range over which these calibrations are valid and will yield reliable results using the following method.
We first calculate the running median O/H and uncertainty on this median in bins in O/H that are 0.2~dex wide.
The uncertainty on the median includes both the measurement uncertainties for individual objects as well as 
 sample variance by bootstrap resampling, the latter of which especially affects the low and high metallicity regimes where are sample is sparse.
We use the uncertainty on the median O/H as a function of median O/H to determine the range of metallicities over which
 our sample can constrain O/H with suitable precision, based on the number of galaxies at each metallicity and the magnitude of the
 uncertainty on the individual O/H determinations.
We define the valid range of the calibrations as the range of metallicities over which the median O/H uncertainty is $<0.1$~dex.
We find that our new calibrations are valid over the range $7.4<12+\mbox{log(O/H)}<8.3$.
Our sample provides the highest precision at 12+log(O/H$)=7.7-8.2$ where the median O/H uncertainty is $<0.05$~dex, while
 at $7.4<12+\mbox{log(O/H)}<7.7$ and $8.2<12+\mbox{log(O/H)}<8.3$ the typical median O/H uncertainty is 0.07~dex.
At 12+log(O/H$)<7.4$ and 12+log(O/H$)>8.3$,
 it is primarily the small number of sources in the current sample that drives an unacceptably large uncertainty.
One clear path to improve the sample at low metallicities is deeper {\it JWST}/NIRSpec spectroscopy of faint low-mass targets \citep[e.g.,][]{las2023}.

\begin{table}
 \centering
\caption{Best-fit coefficients for the high-redshift strong-line metallicity calibrations (equation~\ref{eq:cal} and  Fig.~\ref{fig:cals}).
 }\label{tab:cal}
 \setlength{\tabcolsep}{2.5pt}
 \renewcommand{\arraystretch}{1.5}
 \begin{tabular}{ l r r r r r r  }
   \hline\hline
   $R$\tablenotemark{a}  &
   $N_{\text{gal}}$\tablenotemark{b} &
   c$_0$  &
   c$_1$  &
   c$_2$  &
  $\sigma_{R,\mathrm{fit}}$\tablenotemark{c}  & $\sigma_{R,\mathrm{int}}$\tablenotemark{d}  \\
   \hline\hline
O3  &  46  &  $0.834$  &  $-0.072$  &  $-0.453$  &  0.02  &  0.09    \\  
O2  &  39  &  $0.067$  &  $1.069$  &  ---  &  0.07  &  0.22    \\  
R23  &  43  &  $1.017$  &  $0.026$  &  $-0.331$  &  0.02  &  0.06    \\  
O32  &  39  &  $0.723$  &  $-1.153$  &  ---  &  0.09  &  0.29    \\  
Ne3O2 &  31  &  $-0.386$  &  $-0.998$  &  ---  &  0.07  &  0.24    \\  
   \hline\hline
 \end{tabular}
\tablenotetext{a}{Strong-line ratio.}
\tablenotetext{b}{Number of galaxies with detections of $R$ included in the fit.}
\tablenotetext{c}{Estimate of the formal uncertainty in $R$ at fixed O/H of the best-fit calibration.}
\tablenotetext{d}{Intrinsic scatter in $R$ at fixed O/H about the best-fit calibration, after accounting for measurement uncertainties.}
\end{table}

We calculate the intrinsic scatter in line ratio at fixed O/H by computing a $\chi^2$ statistic that includes the measurement uncertainty in both parameters as well as an intrinsic scatter term:
\begin{equation}
\chi^2 = \sum \frac{(F_R(\mathrm{O/H})-R_{\mathrm{obs}})^2}{(\sigma_{R,\mathrm{obs}}^2 + (\dot{F}_R\sigma_{\mathrm{O/H,obs}})^2 + \sigma_{R,\mathrm{int}}^2)}
\end{equation}
where $F_R$ is the best-fit calibration for ratio $R$, $R_{\mathrm{obs}}$ is the observed line ratio, $\sigma_{R,\mathrm{obs}}$ is the measurement uncertainty on $R$, $\dot{F}_R$ is the derivative of $F_R$ evaluated at O/H of the source, $\sigma_{\mathrm{O/H,obs}}$ is the measurement uncertainty on O/H, $\sigma_{R,\mathrm{int}}$ is the intrinsic scatter term, and the sum is evaluated over all objects with a detection for $R$.
We then vary the intrinsic scatter term to find the value of $\sigma_{R,\mathrm{int}}$ for which the reduced $\chi^2$ is equal to unity.
The inferred intrinsic scatters are reported in Table~\ref{tab:cal}, ranging from 0.06 to 0.29~dex.
These values are generally similar to what has been found for $z\sim0$ galaxies and H\ii\ regions for which O3 and R23 also show smaller intrinsic scatter than O32 and Ne3O2 at fixed O/H \citep[e.g.,][]{mai2008,cur2020}, though this is at least partly due to the fact that O3 and R23 span a smaller dynamic range than O32 and Ne3O2 in our sample.
For the linear fits of O2, O32, and Ne3O2, we can convert the intrinsic scatter in $R$ to an intrinsic scatter in O/H at fixed strong-line ratio of $\sigma_{\mathrm{O/H,int}}=0.20$, 0.25, and 0.24~dex.

\section{Discussion}\label{sec:discussion}

\subsection{Comparison to literature calibrations}

\begin{figure*}
\centering
\includegraphics[width=\textwidth]{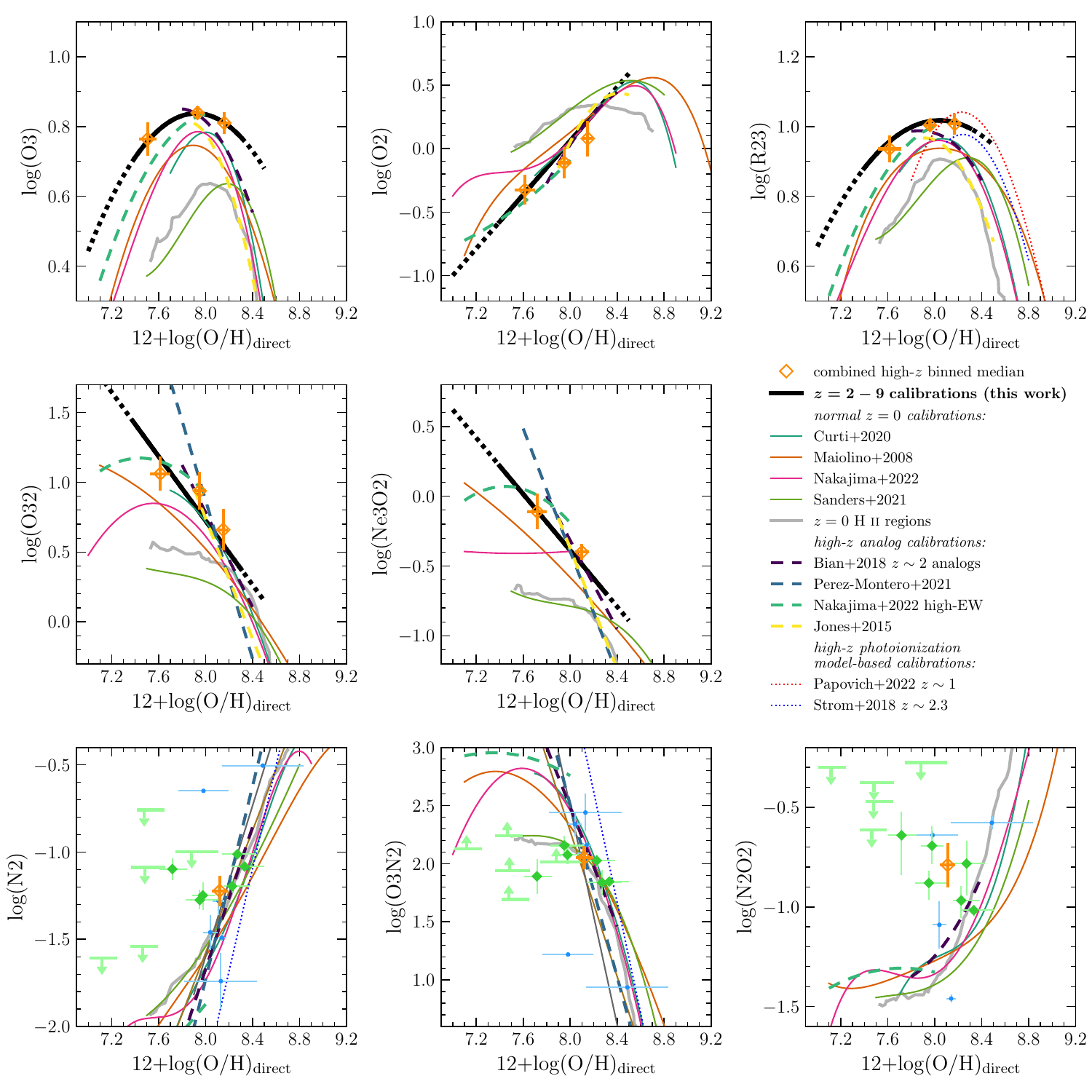}
\caption{Comparison of the best-fit high-redshift calibrations (black lines, as in Fig.~\ref{fig:cals}) and high-redshift binned medians (orange diamonds) to a selection of strong-line calibrations from the literature including those calibrated to ``normal'' $z\sim0$ star-forming galaxies and H\ii\ regions \citep{cur2020,mai2008,nak2022,san2021}; extreme local galaxies that are analogs of high-redshift galaxies \citep{bia2018,per2021,nak2022,jon2015}; and calibrations based on the application of photoionization model fitting to strong-line samples at $z\sim1-3$ \citep{pap2022,str2018}.
For the line ratios involving [N\ii] in the bottom row, we show the full set of high-redshift points since robust calibrations could not be fit with existing data for N2, O3N2, and N2O2.
}\label{fig:compare}
\end{figure*}

We compare the new empirical high-redshift calibrations derived in Sec.~\ref{sec:results} to existing calibrations in the literature based on representative $z\sim0$ sources and extreme local galaxies that are analogs of high-redshift systems, as well as semi-empirical calibrations derived by applying photoionization modeling to high-redshift spectroscopic samples.
Figure~\ref{fig:compare} compares our high-redshift calibrations (black lines) and binned medians (orange diamonds) to the literature strong-line calibrations.

\subsubsection{Normal $z\sim0$ calibrations}

We first compare to ``normal'' local-universe calibrations based on $z=0$ H\ii\ regions and/or $z\sim0$ star-forming galaxies, shown as solid colored lines in Fig.~\ref{fig:compare} \citep{mai2008,cur2017,cur2020,san2021,nak2022}.
Here, the term normal corresponds to calibrations for which the calibrating sample has ISM ionization conditions representative of what is typical at $z\sim0$, defined either by the average properties of nearby H\ii\ regions or by galaxies falling on the $z\sim0$ star-forming main sequence.
The high-redshift calibrations are distinct from normal local calibrations in several ways.
The $z\sim0$ calibrations fail to reach the high O3 and R23 values that are common among the high-redshift sample, and generally have lower O3 and R23 at fixed O/H relative to the high-redshift calibrations.
For the ionization-sensitive ratios, the high-redshift calibrations have higher O32 and Ne3O2 than the local calibrations at fixed O/H.
At fixed O/H, O2 may be slightly lower at high-redshift than at $z\sim0$ although the distinction is less clear than for O3, R23, O32, and Ne3O2.
These offsets demonstrate that H\ii\ regions in high-redshift galaxies are more highly ionized than their local counterparts at fixed metallicity.
This result is consistent with many studies that have concluded that the evolution of line ratio excitation sequences between $z\sim0$ and $z\sim2-3$ arises because high-redshift galaxies have harder ionizing spectra at fixed O/H \citep[e.g.,][]{ste2016,str2018,sha2019,san2020,top2020a,top2020b,jeo2020,cul2021,run2021}.
The difference between the $z\sim0$ and new high-redshift calibrations shows how a different set of ionization conditions results in a change in the form of the metallicity calibrations.
We conclude that normal local-universe calibrations should not be applied to high-redshift samples and will yield biased metallicity inferences if they are used at $z\gtrsim2$.
For O2, O32, and Ne3O2, local calibrations tend to underestimate the metallicity of high-redshift samples by $\sim0.1-0.4$~dex, consistent with what was found in \citet{san2021}.
For O3 and R23 the direction of the bias will depend on whether an object is on the lower or upper branch.

\subsubsection{Local analog calibrations}

The use of calibrations based on extreme local galaxies that have properties analogous to high-redshift galaxies has recently become commonplace among high-redshift metallicity studies employing strong-line methods \citep[e.g.,][]{san2020,san2021,wan2022,mat2022,li2022,nak2023}.
The dashed lines in Figure~\ref{fig:compare} show different empirical local analog calibrations \citep{jon2015,bia2018,per2021,nak2022}.
\citet{jon2015} fit their calibration to \oiiia-detected $z\sim0$ galaxies from SDSS that have high sSFRs due to this selection.
We note also that the low metallicity (12+log(O/H$)\lesssim8.1$) data used by \citet{cur2020} also employed individual \oiiia-detected SDSS galaxies that mostly fall near the $z\sim2$ star-forming main sequence \citep{san2021}, explaining why the \citet{cur2020} calibrations tend to match the local analogs at low metallicity.

We find that the local analog calibrations perform better than the normal $z\sim0$ calibrations in matching the high-redshift sample, consistent with what was found using ground-based \te-metallicities for $\sim20$ galaxies at $z\sim2$ by \citet{san2020}.
One challenge in using local analogs is that their metallicity range is typically limited, thus limiting the usable range of the calibration without relying on extrapolation.
For example, the \citet{nak2022} analog calibrations only cover very low metallicities at 12+log(O/H$)<8.0$.
The \citet{bia2018} sample spans 12+log(O/H$)=7.8-8.4$, failing to reach low enough metallicities to be relevant for many of the $z>6$ or low-mass $z\sim2$ galaxies that now have {\it JWST} spectra.
In contrast, our high-redshift calibrations are valid over 12+log(O/H$)=7.4-8.3$, offering useful calibrations over a wider metallicity range.

The local analog calibrations match our new calibrations relatively well in O32 and Ne3O2 though there is some deviation at 12+log(O/H$)\lesssim7.8$, where the \citet{nak2022} relation is flatter while the \citet{bia2018} relation is steeper.
The disagreement with \citet{nak2022} at low metallicities may be a result of the choice of parameterization,
 as we assumed a linear function for O32 and Ne3O2 while they utilized a second-order polynomial.
The two lowest metallicity galaxies in our sample fall below our best-fit linear O32 calibration potentially indicating a flattening,
 but better statistics in the metal-poor regime are needed to robustly evaluate this possibility.
For O3 and R23, the local analog calibrations appear to fall off more steeply in both the lower and upper branch than the high-redshift calibrations.
The slope of the high-redshift calibrations at the metal-rich (12+log(O/H$)>8.3$) and metal-poor (12+log(O/H$)<7.4$) ends
 remains uncertain due to poor statistics.
However, a more gradual O3 and R23 turnover and falloff than the local analog calibrations
 is evident even in the range where the high-redshift sample is well populated
 ($7.4<12+\mbox{log(O/H)}<8.3$), such that the high-redshift sample has higher O3 and R23 at fixed O/H than most of the local analogs on average.
Increasing the number of high-redshift galaxies with auroral line detections at very low and high metallicities would
 provide a stringent test of the local analog method since our calibrations predict an even larger discrepancy in those regimes.
In any case, our new calibrations, derived directly a large \te-based sample at $z=2-9$, offer a more
 direct route to accurate strong-line metallicities in the early universe than relying on the indirect analog connection.

\subsubsection{Model-based high-redshift calibrations}

\citet{str2018} and \citet{pap2022} constructed semi-empirical calibrations using $z\sim1-3$ star-forming galaxies by first employing photoionization model grids to deriving metallicities of the samples and then fitting strong-line ratio vs.\ metallicity relations to the resulting distributions.
While these model-based calibrations reach high enough R23 values to match the peak of the high-redshift sample used in this work, the turnover point is shifted $0.2-0.3$~dex higher in metallicity (top right panel of Fig.~\ref{fig:compare}).
Likewise, the \citet{str2018} N2 and O3N2 calibrations are also shifted toward higher O/H at fixed line ratio relative to the high-redshift median.
These model-based calibrations do not reproduce metallicities on the empirical \te\ scale.

\subsubsection{Nitrogen-based indicators at high redshift}

The utility of line ratios including [N\ii] to derive metallicities at high redshift has been a subject of debate since it was pointed out that N/O may evolve with redshift at fixed O/H or be less tightly coupled to O/H at high redshifts \cite[e.g.,][]{mas2014,mas2016,ste2014,ste2016,san2016,str2017,str2018,str2022,hay2022,san2023oiia}.
Interestingly, strong rest-UV N\iii] and N\iv] lines recently reported in the spectrum of GN-z11 at $z=10.6$ potentially imply a super-solar N/O ratio despite having $\sim10$\% solar O/H \citep{bun2023,cam2023}.
Despite early {\it JWST} observations more than doubling the existing high-redshift auroral-line sample, [N\ii] is only detected for 12/46 objects with [N\ii] upper-limits dominating at 12+log(O/H$)<7.9$.
The inherent weakness of [N\ii] in metal-poor high-redshift systems by itself suggests that N-based calibrations are far less useful than those based on O and Ne line ratios at $z>2$.
Comparing the median of the 12 [N\ii]-detected sources to the literature calibrations, we find that the existing calibrations have lower N2 at fixed O/H on average than the high-redshift sample.
The median O3N2, on the other hand, matches existing \te-based calibrations well, with the apparent evolution toward higher O3 and N2 at fixed O/H canceling out.
Focusing on N2O2, we find that the [N\ii]-detected high-redshift galaxies have higher N2O2 at fixed O/H than both $z\sim0$ normal and analog calibrations.
Since N2O2 correlates tightly with N/O \citep[e.g.,][]{per2009,str2017}, this offset suggests that the N/O vs.\ O/H relation found at $z\sim0$ does not hold in the same form at high redshift.
A significantly expanded sample of high-redshift galaxies with both \te\ measurements and [N\ii] detections is required to robustly assess the N-based indicators.
Until such a sample is available, metallicity indicators based on [N\ii] should be used with great caution at high redshifts.

\subsection{Applicability of the calibrations over $z=2-9$}\label{sec:zrange}

A question that must be addressed is whether a single set of calibrations can truly yield accurate results over the wide redshift range of our sample, spanning $z=2-9$.
A single set of metallicity calibrations could hold if the typical ionization conditions in H\ii\ regions do not significantly evolve over this redshift interval.
Using a sample of 164 star-forming galaxies at $z=2-9$ from CEERS, in \citet{san2023bpt} we recently showed that galaxies at $z=2.7-6.5$ fall on the same excitation sequence as $z=2.0-2.7$ galaxies in the [O\iii]/H$\beta$ vs.\ [N\ii]/H$\alpha$ and [S\ii]/H$\alpha$ ``BPT'' diagrams and the O32 vs.\ R23 diagram.
This result suggests that, within the constraints offered by the admittedly limited current {\it JWST} spectroscopic samples, ISM ionization conditions do not significantly change between $z\sim2$ and $z\sim6$.
In contrast, clear evolution in these excitation sequences is present between $z\sim0$ and $z\sim2$, demonstrating distinct ionization conditions that manifest as distinct line ratio vs.\ metallicity sequences as shown in this work.

\begin{figure}
\centering
\includegraphics[width=\columnwidth]{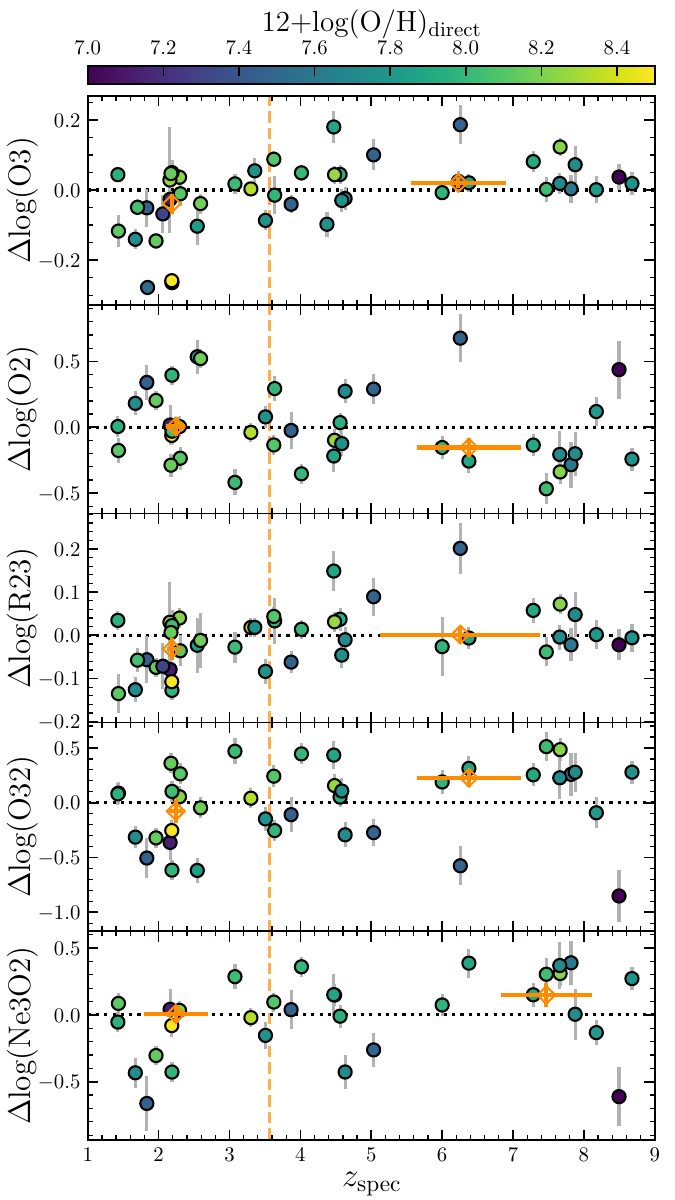}
\caption{Residuals in strong-line ratios relative to the best-fit calibrations (Table~\ref{tab:cal}) vs.\ spectroscopic redshift for the combined high-redshift sample, color-coded by O/H.
The orange diamonds display binned medians after dividing the sample into two redshift bins, split at the median redshift of
 $z_{\mathrm{med}}=3.6$ (orange dashed line).
}\label{fig:res}
\end{figure}

Figure~\ref{fig:res} shows the residuals in the strong-line ratios O3, O2, R23, O32, and Ne3O2
 relative to our best-fit calibrations vs.\ redshift for the combined aurora-line sample, color-coded by O/H.
We find that, within this sample, metallicity is fairly evenly distributed as a function of redshift such that the fit in a particular metallicity range is not dominated by galaxies from the lower or higher end of the redshift range.
We performed a Spearman correlation test on the individual galaxies and used 10,000 Monte Carlo realizations to assess the
 significance of any possible correlations including purturbations according to the uncertainties \citep{cur14},
 calculating Spearman correlation coefficients, $p\mbox{-values}$, and confidence intervals on both quantities.
We find that there are no statistically significant correlations present among the individual targets for any of these line ratios,
 where a significant correlation is here defined as the case where $p\mbox{-value}<0.0027$ with $>3\sigma$ confidence
 (i.e., the hypothesis that the two samples are drawn from independent distributions is rejected confidently at the $>3\sigma$ level).
The only line ratio for which the $p\mbox{-value}$ is less than 0.0027 is O3, with $p\mbox{-value}=0.00058^{+0.0062}_{-0.00037}$
 such that the uncertainty introduced by the measurement errors is sufficiently large that this apparent correlation cannot be trusted
 with high significance.
This result suggests that, within the limits of the current sample, there is not significant redshift evolution relative to our new calibrations.

Nevertheless, there are tentative trends for some line ratios.
While O3 and R23 do not appear to display any significant postive or negative residuals on average as a function of redshift,
 the majority of the $z>7$ points have positive O32 and Ne3O2 residuals and negative O2 residuals.
To better assess these trends, we calculate median values in two bins of redshift after dividing the sample at the median redshift
 of $z_{\mathrm{med}}=3.6$, as well as uncertainties on the medians including measurement uncertainty and sample variance via bootstrap resampling.
We find that O3 and R23 display little evidence for evolution over this redshift range, with median offsets of $<0.05$~dex in line ratio at fixed
 O/H and $\le1.5\sigma$ consistency between the low and high redshift bins.
O32 displays the most significant trend, with the high-redshift bin displaying a median O32 value $2.2\sigma$ higher than the low-redshift bin.
These trends hint at a possible evolution toward higher ionization states with increasing redshift, particularly at $z>7$,
 that would result in a need to revise the strong-line calibrations at such high redshifts.
However, the current sample is clearly not large enough to evaluate such evolution with sufficient confidence.
This question must be revisited when larger samples with higher S/N are available from future observations with {\it JWST}.

\subsection{Appropriate uses of the calibrations}

To ensure robust metallicity inferences, it is important that strong-line calibrations like the ones constructed in this work are used appropriately.
First, strong-line calibrations should ideally only be used on samples that fall within the same line ratio and metallicity range as the calibration sample, otherwise an uncertain extrapolation is required.
For our new high-redshift calibrations, the valid range is 12+log(O/H$)=7.4-8.3$.
Second, it is clear in Fig.~\ref{fig:cals} and in Table~\ref{tab:cal} that the intrinsic scatter of these relations is large.
This is generally true of all strong-line calibrations, whether local or high redshift, because the correlations between nebular metallicity and properties including ionization parameter, density, ionizing spectral shape, and N/O have significant scatter \citep[e.g.,][]{per2009,san2016,per2014}.
Consequently, metallicity derived via the strong-line method for a single object necessarily carries a large uncertainty.
However, determining the average metallicity across a sample, potentially in multiple bins, can achieve a high degree of accuracy, where the uncertainty due to intrinsic scatter reduces by $\sqrt{N}$ for a sample of $N$ objects.
Metallicities derived for single galaxies using our new calibrations should thus include uncertainty due to the intrinsic scatter.
Ideally, studies of the MZR and FMR in the high-redshift universe should utilize sufficiently large samples to statistically reduce the effects of this intrinsic scatter to obtain accurate mean relations.
Finally, we caution against the use of these calibrations (or normal $z\sim0$ calibrations) with samples at intermediate redshifts (i.e., $z=0.5-1.5$), which appear to have ionization conditions distinct from those at $z\sim0$ but less extreme than those at $z\sim2$ \citep[e.g.,][]{sha2019,hir2021}.

\subsection{Areas of improvement for high-redshift calibrations}\label{sec:areas}

While the new calibrations presented here represent a major step forward for high-redshift metallicity determinations, there remain clear directions to improve these calibrations with additional observations.
The metallicity ranges encompassing the lower and upper O3 and R23 branches at 12+log(O/H$)\lesssim7.5$ and 12+log(O/H$)\gtrsim8.3$ are not well populated compared to the O3 and R23 peak where the majority of the sample lies.
The O3 and R23 peak is the region in which [O\iii]$\lambda$4364 will be brightest relative to Balmer lines like H$\beta$ and H$\alpha$, such that this is the easiest metallicity range in which to detect [O\iii]$\lambda$4364.
Improving statistics for [O\iii]$\lambda$4364 at both low and high metallicities requires deeper spectroscopy than the CEERS/NIRSpec medium grating observations.
Since this program had on-source integration times of $\sim1$~hour per grating, significant improvement in the limiting line flux will be achieved in {\it JWST}/NIRSpec programs featuring several-hour integrations.
There is additional promise for obtaining more \te\ constraints at high metallicity (12+log(O/H$)\gtrsim8.4=0.5~Z_{\odot}$) using the low-ionization \oiia\ auroral lines that should be detectable at higher metallicities and lower \te\ than \oiiia.

The paucity of [N\ii] detections is another clear weakness of the current high-redshift auroral-line sample, preventing the robust assessment and formulation of calibrations for N-based line ratios that are among the most common metallicity indicators used for local-universe samples.
Deeper spectroscopy is again the solution, where achieving reasonable completeness in [N\ii] for a sample similar to the one used in this study must have sufficient sensitivity to detect lines $\gtrsim$30 times weaker than H$\alpha$ \citep[see also][]{san2023bpt,sha2023b}.

A final and significant outstanding problem is whether typical main-sequence galaxies at these redshifts follow the same calibration relations as the objects with detected auroral lines in the current sample.
Selecting samples based on the detection of very faint auroral emission lines necessarily introduces a bias, typically toward high-SFR and high-[O\iii] equivalent width sources.
\citet{san2020} demonstrated that this is indeed the case for the ground-based $z\sim2-3$ auroral-line sample.
We find that the {\it JWST} auroral-line emitters at $z\sim2-6$ likewise fall above the mean star-forming main sequence at these redshifts \citep[Fig.~\ref{fig:sfrmstar};][]{spe2014}.
At $z>6$, more {\it JWST} spectroscopy and imaging is required to robustly characterize the typical star-forming population before we can robustly asses how representative (or not) the \te\ sample is.
{\it JWST} shows great promise to address all of these shortcomings in the next few years.

\section{Summary and Conclusions}\label{sec:summary}

We report detections of the temperature-sensitive [O\iii]$\lambda4364$ auroral emission line for 16 galaxies at $z=2.1-8.7$ from the CEERS survey, measured from medium-resolution {\it JWST}/NIRSpec observations.
The \oiia\ auroral emission line was also detected for two of these sources, the first high-redshift galaxies with constraints on \te\ in both the low- and high-ionization zones.
We combine the CEERS sample with 9 galaxies at $z=4-9$ from the literature with auroral-line detections from {\it JWST}/NIRSpec and 21 objects with auroral-line detections from ground-based spectroscopy at $z=1-4$.
The combined high-redshift auroral-line sample comprises 46 star-forming galaxies at $z=1.4-8.7$, more than doubling the sample size from ground-based observations over the past decade.
We calculate \te\ and direct-method oxygen abundances for this sample, and construct the first \te-based empirical strong-line metallicity calibrations based purely on high-redshift galaxies (Fig.~\ref{fig:cals}).
These calibrations, presented in Table~\ref{tab:cal}, are valid over the metallicity range 12+log(O/H$)=7.4-8.3$.

Our new calibrations, derived directly from observations of high-redshift sources, represent a significant step forward in our ability to derive accurate metallicities in the early universe.
Studies of the MZR and FMR at high redshifts no longer need to rely on local-universe metallicity calibrations or the indirect approach in which extreme local galaxies are assumed to be analogs of high-redshift systems.
These measurements also provide important empirical tests for any theoretical photoionization model-based methods of deriving metallicities at high redshift.

We compared these calibrations to strong-line calibrations from the literature, finding that the high-redshift calibrations have higher O3, R23, O32, and Ne3O2 line ratios at fixed O/H relative to normal $z\sim0$ calibrations.
This redshift evolution of strong-line calibration functions is driven by evolving ISM ionization conditions between $z\sim0$ and $z\ge2$ identified in studies of star-forming galaxies at $z\sim2-3$ \citep[e.g.,][]{ste2016,sha2019,top2020a}.
Local analog calibrations display much better agreement with the high-redshift data for these line ratios and in particular reach the high O3 and R23 values that normal $z\sim0$ calibrations fail to reproduce, but still do not display consistent agreement with the high-redshift calibrations across all line ratios and regions of parameter space,
falling off more steeply from the maximum O3 and R23 value than the high-redshift calibrations.

The current high-redshift \te\ sample features only 12 detections of the [N\ii]$\lambda$6585 emission line used in some of the most commonly-used metallicity indicators at $z\sim0$, including the N2 and O3N2 ratios.
Consequently, calibrations of N-based indicators cannot yet be robustly assessed at high redshift.
The current sample is also lacking good statistics at both very low (12+log(O/H$)\lesssim7.4$) and high (12+log(O/H$)\gtrsim8.3$) metallicities.
Deep spectroscopy with several hours of integration per pointing with {\it JWST}/NIRSpec promises to improve both of these shortcomings by detecting fainter \oiiia\ lines at low metallicity and low-ionization \oiia\ lines at high metallicity, while also increasing the detection rate of [N\ii] which has proven to be very faint in $z>4$ systems.

The new high-redshift metallicity calibrations presented in this work will yield an immediate improvement to strong-line metallicities in existing and future $z>2$ spectroscopic samples.
They are applicable over a redshift range spanning from Cosmic Noon into the Epoch of Reionization, leading to a more accurate characterization of the evolution of metallicity scaling relations.
These improved constraints will in turn provide insight into the nature of galaxy growth, feedback, and baryon cycling in the early universe.

\begin{acknowledgments}
We would like to thank the entire CEERS team for designing and executing the Early Release Science observations used in this work, most especially for designing the NIRSpec MSA observations.
This work is based on observations made with the NASA/
ESA/CSA James Webb Space Telescope. The data were
obtained from the Mikulski Archive for Space Telescopes at
the Space Telescope Science Institute, which is operated by the
Association of Universities for Research in Astronomy, Inc.,
under NASA contract NAS5-03127 for {\it JWST}. 
The specific observations analyzed can be accessed
via \href{https://archive.stsci.edu/doi/resolve/resolve.html?doi=10.17909/z7p0-8481}{DOI: 10.17909/z7p0-8481}.
Support for this work was provided through the NASA Hubble Fellowship
grant \#HST-HF2-51469.001-A awarded by the Space Telescope
Science Institute, which is operated by the Association of Universities
for Research in Astronomy, Incorporated, under NASA contract NAS5-26555.
We also acknowledge support from NASA grant {\it JWST}-GO-01914.
The Cosmic Dawn Center is funded by the Danish National Research Foundation (DNRF) under grant \#140.  Cloud-based data processing and file storage for this work is provided by the AWS Cloud Credits for Research program.
\end{acknowledgments}

\vspace{5mm}
\facilities{{\it JWST}(NIRSpec)}

\appendix

\section{Line flux measurements for the CEERS auroral-line sample}\label{app:lines}

The measured line fluxes for the CEERS auroral-line sample are reported in Table~\ref{tab:lines}.
If a line was not detected at $>3\sigma$ significance, a $3\sigma$ upper limit is instead given.
Lines with no value reported were not covered in the spectral range of the observations.

\begin{sidewaystable*}
 \caption{Measured rest-optical emission-line fluxes for the CEERS auroral-line sample, in units of $10^{-18}$~erg~s$^{-1}$~cm$^{-2}$.
 }\label{tab:lines}
 \movetableright=-5cm
 \setlength{\tabcolsep}{2.5pt}
 \renewcommand{\arraystretch}{1.5}
 \resizebox{\textheight}{!}{
 \begin{tabular}{ r r r r r r r r r r r r r r r r }
   \hline\hline
  ID\tablenotemark{a}  &  $[$O~\textsc{ii}$]$$\lambda$3728\tablenotemark{b}  &  $[$Ne~\textsc{iii}$]$$\lambda$3870  &  H$\delta$  &  H$\gamma$  &  $[$O~\textsc{iii}$]$$\lambda$4364  &  H$\beta$  &  $[$O~\textsc{iii}$]$$\lambda$4960  &  $[$O~\textsc{iii}$]$$\lambda$5008  &  H$\alpha$  &  $[$N~\textsc{ii}$]$$\lambda$6585  &  $[$S~\textsc{ii}$]$$\lambda$6718  &  $[$S~\textsc{ii}$]$$\lambda$6733  &  $[$O~\textsc{ii}$]$$\lambda$7322,32\tablenotemark{c}  \\  
   \hline\hline
  1019  &  $3.93$$\pm$$0.38$  &  $5.04$$\pm$$0.33$  &  $2.50$$\pm$$0.33$  &  $5.43$$\pm$$0.40$  &  $1.81$$\pm$$0.35$  &  $10.17$$\pm$$0.55$  &  $23.00$$\pm$$0.52$  &  $71.30$$\pm$$0.63$  &  ---  &  ---  &  ---  &  ---  &  ---  \\
  1149  &  $1.18$$\pm$$0.08$  &  $0.55$$\pm$$0.07$  &  $0.23$$\pm$$0.08$  &  $0.77$$\pm$$0.08$  &  $0.23$$\pm$$0.07$  &  $1.52$$\pm$$0.09$  &  $3.39$$\pm$$0.10$  &  $10.61$$\pm$$0.09$  &  ---  &  ---  &  ---  &  ---  &  ---  \\
  1027  &  $3.17$$\pm$$1.02$  &  $7.93$$\pm$$0.82$  &  $2.77$$\pm$$0.66$  &  $7.18$$\pm$$0.68$  &  $2.97$$\pm$$0.49$  &  $17.00$$\pm$$1.03$  &  $36.32$$\pm$$0.86$  &  $109.70$$\pm$$0.86$  &  ---  &  ---  &  ---  &  ---  &  ---  \\
  698  &  $5.35$$\pm$$0.92$  &  $4.10$$\pm$$0.62$  &  $3.33$$\pm$$0.66$  &  $5.90$$\pm$$0.64$  &  $1.27$$\pm$$0.53$  &  $12.85$$\pm$$0.72$  &  $28.22$$\pm$$0.82$  &  $87.84$$\pm$$0.81$  &  ---  &  ---  &  ---  &  ---  &  ---  \\
  792  &  $2.04$$\pm$$0.27$  &  $<$$0.67$  &  $1.31$$\pm$$0.24$  &  $0.88$$\pm$$0.18$  &  $0.63$$\pm$$0.15$  &  $1.33$$\pm$$0.16$  &  $3.78$$\pm$$0.12$  &  $11.46$$\pm$$0.13$  &  $3.96$$\pm$$0.16$  &  $<$$0.69$  &  $<$$0.49$  &  $<$$0.52$  &  ---  \\
  397  &  $4.08$$\pm$$0.20$  &  $2.11$$\pm$$0.17$  &  $1.16$$\pm$$0.18$  &  $2.48$$\pm$$0.20$  &  $0.64$$\pm$$0.13$  &  $5.59$$\pm$$0.13$  &  $11.46$$\pm$$0.13$  &  $37.91$$\pm$$0.13$  &  $16.43$$\pm$$0.15$  &  $0.93$$\pm$$0.17$  &  $<$$0.37$  &  $<$$0.37$  &  $<$$0.86$  \\
  1536  &  $1.78$$\pm$$0.34$  &  $1.31$$\pm$$0.21$  &  $1.24$$\pm$$0.26$  &  $1.43$$\pm$$0.20$  &  $0.89$$\pm$$0.21$  &  $2.79$$\pm$$0.25$  &  $7.48$$\pm$$0.24$  &  $19.80$$\pm$$0.23$  &  $9.16$$\pm$$0.13$  &  $<$$0.75$  &  $<$$0.39$  &  $<$$0.45$  &  $<$$0.81$  \\
  1477  &  $11.68$$\pm$$0.93$  &  $3.47$$\pm$$0.76$  &  $3.30$$\pm$$0.52$  &  $4.98$$\pm$$0.46$  &  $1.97$$\pm$$0.41$  &  $11.64$$\pm$$0.57$  &  $25.98$$\pm$$0.78$  &  $73.15$$\pm$$0.77$  &  $39.27$$\pm$$0.45$  &  $3.14$$\pm$$0.42$  &  $1.15$$\pm$$0.35$  &  $<$$2.79$  &  $<$$2.02$  \\
  1746  &  $3.93$$\pm$$0.20$  &  $1.77$$\pm$$0.17$  &  $1.09$$\pm$$0.12$  &  $1.76$$\pm$$0.12$  &  $0.57$$\pm$$0.15$  &  $3.51$$\pm$$0.13$  &  $7.90$$\pm$$0.14$  &  $26.69$$\pm$$0.14$  &  $9.78$$\pm$$0.08$  &  $0.52$$\pm$$0.07$  &  $0.32$$\pm$$0.07$  &  $0.34$$\pm$$0.08$  &  $<$$0.50$  \\
  1665  &  $40.55$$\pm$$1.03$  &  $12.80$$\pm$$1.07$  &  $7.06$$\pm$$0.88$  &  $11.04$$\pm$$0.78$  &  $1.75$$\pm$$0.53$  &  $26.79$$\pm$$0.77$  &  $59.30$$\pm$$0.82$  &  $182.90$$\pm$$0.97$  &  $96.92$$\pm$$0.62$  &  $9.44$$\pm$$0.56$  &  $4.05$$\pm$$0.44$  &  $3.33$$\pm$$0.55$  &  $<$$2.35$  \\
  1559  &  $1.14$$\pm$$0.23$  &  $0.87$$\pm$$0.20$  &  $0.75$$\pm$$0.19$  &  $1.51$$\pm$$0.20$  &  $0.64$$\pm$$0.19$  &  $2.15$$\pm$$0.20$  &  $6.10$$\pm$$0.18$  &  $22.37$$\pm$$0.19$  &  $6.00$$\pm$$0.16$  &  $<$$0.60$  &  $<$$0.59$  &  $<$$0.76$  &  $<$$0.98$  \\
  1651  &  ---  &  ---  &  $<$$0.68$  &  $1.80$$\pm$$0.19$  &  $0.45$$\pm$$0.17$  &  $3.64$$\pm$$0.23$  &  $6.92$$\pm$$0.14$  &  $19.53$$\pm$$0.18$  &  ---  &  ---  &  $<$$0.50$  &  $<$$0.49$  &  $<$$1.04$  \\
  11728  &  $1.58$$\pm$$0.41$  &  $2.46$$\pm$$0.24$  &  $1.16$$\pm$$0.23$  &  $2.86$$\pm$$0.18$  &  $1.00$$\pm$$0.19$  &  $6.03$$\pm$$0.19$  &  $12.53$$\pm$$0.17$  &  $30.81$$\pm$$0.16$  &  $16.82$$\pm$$0.12$  &  $<$$0.49$  &  $<$$0.34$  &  $<$$0.68$  &  $<$$0.60$  \\
  11088  &  $44.25$$\pm$$0.59$  &  $8.36$$\pm$$0.43$  &  $5.90$$\pm$$0.71$  &  $11.58$$\pm$$0.39$  &  $1.51$$\pm$$0.52$  &  $26.32$$\pm$$0.40$  &  $53.71$$\pm$$0.29$  &  $159.30$$\pm$$0.27$  &  $103.60$$\pm$$0.43$  &  $8.59$$\pm$$0.30$  &  $5.65$$\pm$$0.50$  &  $4.51$$\pm$$0.41$  &  $1.91$$\pm$$0.29$  \\
  3788  &  $75.80$$\pm$$1.06$  &  $20.15$$\pm$$0.97$  &  $9.64$$\pm$$0.94$  &  $18.21$$\pm$$0.99$  &  $3.01$$\pm$$0.63$  &  $39.25$$\pm$$1.03$  &  $89.17$$\pm$$1.15$  &  $267.80$$\pm$$0.98$  &  $145.10$$\pm$$0.75$  &  $9.24$$\pm$$0.82$  &  $8.57$$\pm$$0.68$  &  $5.58$$\pm$$0.60$  &  $2.57$$\pm$$0.77$  \\
  3537  &  $0.87$$\pm$$0.26$  &  $3.09$$\pm$$0.19$  &  $1.83$$\pm$$0.18$  &  $3.43$$\pm$$0.15$  &  $1.10$$\pm$$0.20$  &  $8.32$$\pm$$0.15$  &  $10.61$$\pm$$0.19$  &  $28.32$$\pm$$0.20$  &  $30.19$$\pm$$0.18$  &  $<$$0.75$  &  $<$$0.57$  &  $<$$0.57$  &  $<$$0.88$  \\
   \hline\hline
 \end{tabular}}
 \tablenotetext{a}{CEERS ID number.}
 \tablenotetext{b}{Sum of [O\ii]$\lambda$3727 and [O\ii]$\lambda$3730.}
 \tablenotetext{c}{Sum of [O\ii]$\lambda$7322 and [O\ii]$\lambda$7332.}
\end{sidewaystable*}

\bibliography{highzte}{}
\bibliographystyle{aasjournal}

\end{document}